\documentclass[10pt,journal]{IEEEtran}
\ifCLASSINFOpdf
\else
\fi
\usepackage{cite}
\usepackage{graphicx}
\usepackage[warn]{textcomp}
\usepackage{amssymb}
\usepackage{amsmath}
\usepackage{pifont}
\usepackage{multicol}
\usepackage{algorithm}
\usepackage{algorithmicx}
\usepackage{algpseudocode}
\usepackage{float}
\usepackage{setspace}
\usepackage{amsthm}
\usepackage{tikz}
\usepackage{standalone}
\usepackage{inputenc}
\usepackage[T1]{fontenc}
\usepackage{mathtext}
\usepackage{textcomp}
\usepackage[english]{babel}
\usepackage[font=small,labelsep=period]{caption}
\usepackage{xcolor,colortbl}
\usepackage{subcaption} 
\usepackage{balance}

\newlength\figureheight 
\newlength\figurewidth 
\algnewcommand\algorithmicinput{\textbf{Input:}}
\algnewcommand\Input{\item[\algorithmicinput]}

\definecolor{gray}{rgb}{0.86, 0.86, 0.86}
\newenvironment{rcases}
  {\left.\begin{aligned}}
  {\end{aligned}\right\rbrace}
  \newcommand{\cmark}{\ding{51}}%
\newcommand{\xmark}{\ding{55}}%
\algnewcommand\algorithmicforeach{\textbf{for each:}}
\algnewcommand\ForEach{\item[ \algorithmicforeach]}
\let\oldReturn\Return
\renewcommand{\Return}{\State\oldReturn}
\newtheoremstyle{component}{}{}{}{}{\bfseries}{.}{.5em}{\thmnote{#3}#1}
    \theoremstyle{component}
    
\algnewcommand\algorithmicoutput{\textbf{Output:}}
\algnewcommand\Output{\item[\algorithmicoutput]}
\newlength\myindent
\makeatletter
\renewcommand{\fnum@figure}{Fig. \thefigure}
\makeatother

\makeatletter
\algrenewcommand\ALG@beginalgorithmic{\small}
\makeatother
\begin{document}
%
\title{Hiring Doctors in E-Healthcare With Zero Budget}
%
%
%

\author{Vikash~Kumar~Singh,
        Sajal~Mukhopadhyay, Rantu Das
\thanks{V. K. Singh is with the Department
of Computer Science and Enginnering, National Institute of Technology, Durgapur,
WB, 713209 India e-mail: (vikas.1688@gmail.com).}
\thanks{S. Mukhopadhyay is with the Department
of Computer Science and Enginnering, National Institute of Technology, Durgapur,
WB, 713209 India e-mail: (sajmure@gmail.com).}
\thanks{R. Das is with the Department
of Computer Science and Enginnering, National Institute of Technology, Durgapur,
WB, 713209 India e-mail: (rantudas247@gmail.com).}}
\maketitle

\begin{abstract}
The doctors (or expert consultants) are the critical resources on which the success of
critical medical cases are heavily dependent. With the emerging technologies (such as video conferencing, smartphone, etc.) this is no longer a dream but a fact, that for critical medical cases in a hospital, expert consultants from around the world could be hired, who may be present physically or virtually. Earlier, this interesting situation of taking the expert consultancies from outside the hospital had been studied, but under monetary perspective. In this 
paper, for the first time, to the best of our knowledge, we investigate the situation, where the \emph{b}elow \emph{i}ncome \emph{g}roup (BIG) people of the society may be served efficiently through the expert consultancy by the renowned doctors from outside of the hospital under \emph{zero} budget. This will help us saving many lives which will fulfil the present day need of biomedical research. We propose three mechanisms: \emph{Ran}dom \emph{p}ick-\emph{a}ssign \emph{m}echanism (RanPAM), \emph{T}ruthful \emph{o}ptimal \emph{a}llocation \emph{m}echanism (TOAM), and \emph{T}ruthful \emph{o}ptimal \emph{a}llocation \emph{m}echanism for in\emph{com}plete \emph{p}references (TOAM-IComP) to allocate the doctor to the patient. With theoretical analysis, we demonstrate that the TOAM is \emph{strategy-proof}, and exhibits a \emph{unique core} property. The mechanisms are also validated with exhaustive experiments. 
\end{abstract}


%
\IEEEpeerreviewmaketitle

\section{Introduction}
\begin{figure}[H]
\centering
 \includegraphics[scale=0.55]{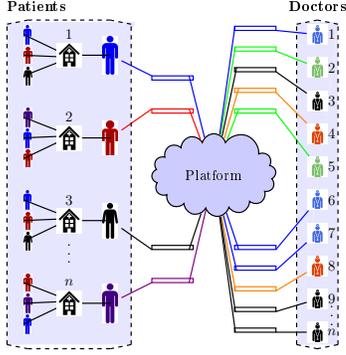}
 \caption{System model}
 \label{fig:1}
\end{figure}

\IEEEPARstart{C}{onsidering} the diverse potential application areas of crowdsourcing \cite{Zhao2014CCC, Yang2012MCN} especially of participatory sensing (PS) \cite{Lee2010PMC, Alshurafa_2015, Kau_bhi}, e-healthcare system, which is emerging as one of the most upcoming technologies to provide an efficient and automated healthcare infrastructure, can employ the crowdsourcing technology to enhance the services provided in that environment. Healthcare consultation (as for example consultation from physician, paediatricians, plastic and cosmetic surgeons, etc.) is said to be the crux of medical  unit and operation suite. Earlier there had been some works to schedule the internal staffs of a medical unit (may be very large)
for working in the operation theatre or controlling the outdoor
units. Most of the literature works on scheduling the internal staffs (inside the hospital) of the medical unit have been devoted to hospital nurses \cite{Berrada_1996, Pierskalla_1994, Weil_1995, Worthington_1988} and the physicians (or doctors) \cite{Carter_2001, Vassilacopoulos_1985}. The doctors are considered as the scarce resource in the medical unit. Scheduling doctors inside a hospital during the critical operations is a challenging task. In \cite{Carter_2001, Vassilacopoulos_1985, Beaulieu_2000, Wang_2007} the different methods of scheduling a physician in an emergency case (may be critical operation) are discussed and presented. Several companies have developed some physician scheduling softwares \cite{Software_1, Software_2, Software_3, Software_4} that will schedule the physicians inside the hospital.  However, it is observed that, with the unprecedented
growth of the communication media (specially Internet), it may
be a common phenomena to hire expert medical staffs (especially
doctors) for a critical operation from outside of the medical unit
where the operation is taking place. This event of hiring an external expert is a special case of crowdsourcing \cite{Zhao2014CCC, Yang2012MCN}  and participatory sensing \cite{Lee2010PMC, TLHP_INFO_2014}. In our future references, hospital, medical unit, organizations will be used interchangeably.\\ 
\indent In the past, operation theatre (OT) planning and scheduling problem \cite{cardoen_b_2010, Magerlein_JM_1978, Blake_J_1997} have been a major area of interest for researchers from various fields and is still an active area of research.   
However, one scenario that may be considered as a research area in healthcare is, say; during an
operation, how to hire a well qualified personnel, including doctors, for a consultation. The challenges come from the following issues: (1) which doctors can be hired? (2) How to motivate the doctors to take part in the system as they may be very busy? (3) If the incentives are provided, how much can be offered? (4) If some renowned doctors are made themselves available for social work, how to grab the situation so that poorest people may be served efficiently to save their valuable life?  In our recent work \cite{Vikash2015EAS}, we have endeavoured solving the problem of hiring one or more expert(s) from outside of the hospital for a critical operations answering some issues related to the challenges mentioned in points 1, 2, and 3. In \cite{Vikash2015EAS} incentives were provided to motivate the doctors for their participation. However, it may be the case, that due to some social awareness, some doctors may impart some social services to the downtrodden community. This situation is mentioned in point 4. In this paper, to the best of our knowledge, first time we have tried to address the practical situation discussed in point 4 above in a game theoretic settings with the robust concept \textbf{m}echanism \textbf{d}esign \textbf{w}ithout \textbf{m}oney (MDWM) or under \emph{zero} budget environment \cite{Gale_AMM_1962, Shapley_2013, NNis_Pre_2007, Roughgarden_T_2013}.\\
\indent In our paper, we have considered \emph{n} hospitals as shown in Fig. \ref{fig:1}. In each hospital several patients for a particular category (say for example gastric ulcer) are admitted who need expert consultation. The patients are of different income groups. Some of the patients may not be able to bear the cost of hiring expert consultation from outside the hospital. However, it may be the case that, due to some social awareness, several doctors throughout the world, may be available freely once in a while (e.g. once in a week). 
We have assumed that, at a particular time more than \emph{n} number of doctors are available. They give their willingness to participate in the consultation process to some \emph{third party} (\emph{platform}). The \emph{third party} selects \emph{n} doctors out of all doctors available based on the quality of the doctors. Now the question is how to use the expert

\noindent consultations that are available freely. In this situation, each hospital selects one patient based on their income group (whoever has lowest income), who will be considered for free consultation shown on the left side in Fig. \ref{fig:1}. Thus, we have \emph{n} patients available for \emph{n} doctors for a particular category (say for paediatric). In this paper, we have first time proposed mechanisms motivated by \cite{Gale_AMM_1962, Shapley_2013, T.roughgarden_2016, Manlove_2016} to allocate the doctors to the patients so that they will be \emph{happy}. By \emph{happy} we mean that each of the participating patients gets their best possible doctor from the available doctors at the time of allocation.\\
The main contributions of this paper are:\\
$\bullet$ First time the problem of hiring doctors is cast as a without money problem or in \emph{zero} budget environment.\\
$\bullet$ The \emph{truthful} mechanisms are proposed for allocating the doctors to the patients.\\
$\bullet$ The simulations are performed for comparing our schemes with a carefully designed benchmark scheme.\\
\\
\indent The remainder of the paper is organized as follows: In section II we describe the system model and formulate the problem as MDWM. In section III, we present two mechanisms: \emph{\textbf{Ran}dom \textbf{p}ick-\textbf{a}ssign \textbf{m}echanism} (RanPAM) and \emph{\textbf{T}ruthful \textbf{o}ptimal \textbf{a}llocation \textbf{m}echanism} (TOAM). Section IV presents the more general set-up in matching market and the mechanism to tackle the more general set-up namely \emph{\textbf{T}ruthful \textbf{o}ptimal \textbf{a}llocation \textbf{m}echanism for \textbf{i}n\textbf{com}plete \textbf{p}references} (TOAM-IComP). A detailed analysis of the experimental results is carried out in section V. Finally, we present a summary of our work and highlight some future directions in section VI.
\section{System model and problem formulation}
We consider the scenario where \emph{n} distinct expert consultants providing their consultancy to \emph{n} different Below Income Group (BIG) patients. One hospital provides one BIG patient for the category under consideration. A platform acts as a third party and decides the $patient-doctor$ allocation pair. Here, we assume that each hospital needs exactly one expert consultant and each expert consultant can provide their service to one hospital at a time. The case that each hospital need multiple expert consultants is left as our future work. In our model, expert consultation may be sought for several categories of diseases. The set of such categories is denoted by $x = \{x_1, x_2, \dots, x_k\}$. The set of sets of available expert consultants in \emph{k} different categories is given as $\mathcal{S} = \{\mathcal{S}_1, \mathcal{S}_2, \ldots, \mathcal{S}_k\}$; where $\mathcal{S}_i$ denote the set of available expert consultants for a particular category $i \in \{1, \ldots, k\}$ and is given as $\mathcal{S}_i = \{s_1^{x_i}, s_2^{x_i}, \dots, s_n^{x_i} \}$. The set of sets of available patients in \emph{k} different categories is given as $\mathcal{P} = \{\mathcal{P}_1, \mathcal{P}_2, \ldots, \mathcal{P}_k\}$; where $\mathcal{P}_i$ is the set of available BIG agents for a particular category $i \in \{1, \ldots, k\}$ and is given as $\mathcal{P}_i$ = $\{ p_1^{x_i}, p_{2}^{x_i}, \dots, p_{n}^{x_i}\}$. When $i^{th}$ \emph{category} is mentioned, the index will be assumed as $i \in \{1, \dots, k\}$, otherwise $i \in \{1, \dots, n\}$. Each agent $p_{i}^{x_i} \in \mathcal{P}_i$ has a strict preference ordering over all $s_{i}^{x_i} \in \mathcal{S}_i$. The strict preference ordering of $t^{th}$ agent $p_{t}^{x_i} \in \mathcal{P}_i$ in $i^{th}$ category is denoted by $\succ^{i}_{t}$ over the set $S_i$, where $s_1^{x_i} \succ^i_{t} s_2^{x_i}$ means that in $i^{th}$ category, $t$ prefers $s_1^{x_i}$ to $s_2^{x_i}$. The set of preferences of all agents in \emph{k} different categories is denoted by $\succ = \{\succ^1, \succ^2, \ldots, \succ^k\}$. Where, $\succ^{i}$ is the preference of all the agents in $i^{th}$ category over all the doctors in $\mathcal{S}_i$, represented as $\succ^{i} = \{\succ^{i}_1, \succ^{i}_2, \dots, \succ^{i}_n\}$. Given the preference of the agents, our proposed mechanisms allocates one doctor to one patient. Let us denote such an allocation vector by $\mathcal{A} = \{\mathcal{A}_1, \mathcal{A}_2, \dots, \mathcal{A}_k\}$; where, each allocation vector $\mathcal{A}_i \in \mathcal{A}$ denotes the allocation vector of agents belongs to the $i^{th}$ category denoted as $\mathcal{A}_i = \{a_1^{i}, a_{2}^{i}, \dots, a_{n}^{i}\}$; where, each $a_t^{i} \in \mathcal{A}_i$ is a $(p_{t}^{x_i}, s_{j}^{x_i})$ pair. Initially, one doctor is allocated to one patient randomly without loss of generality. 

\section{Proposed Mechanisms}
In this section, we have developed two algorithms motivated by \cite{Gale_AMM_1962} \cite{Shapley_2013} \cite{Roughgarden_T_2013}. The first one $i.e.$ RanPAM is given as a naive solution of our problem, that will help to understand better, the more robust \textbf{D}ominant \textbf{s}trategy \textbf{i}ncentive \textbf{c}ompatible (DSIC) mechanism $i.e.$ TOAM presented next.
\subsection{Random Pick-Assign Mechanism (RanPAM)} 
To better understand the model first we propose a \emph{randomized algorithm} called RanPAM to assign doctors to the patients in the hospitals. The RanPAM consists of two stage allocation mechanism, namely; \emph{Main}, and \emph{RanPAM allocation}. 
The idea behind the construction of the \emph{Main} is to capture all the \emph{k} categories present in the system. In each iteration of the \emph{for} loop in line 2-5, a call to \emph{RanPAM allocation} is made. In line 6, the final allocation $\mathcal{A}$ is returned.
 Considering the \emph{RanPAM allocation},        
from line 3 it is clear that the algorithm terminates once the list of patients in $x_i$ category becomes empty. In line 4, the \emph{rand()} function returns the index of the randomly selected patient and is stored in variable $j$. In line 5, the $p^*$ data structure holds the patient present at the index returned in line 4. 
\begin{algorithm}[H]
\caption{Main ($\mathcal{S}$, $\mathcal{P}$, $x$, $\succ$)}
\begin{algorithmic}[1]
      \Output $\mathcal{A}$ $\leftarrow$ $\phi$
	\State \textbf{begin}
	\For{each $x_i$ $\in$ $x$}
	  \State $\mathcal{A}_i$ $\leftarrow$ RanPAM allocation($\mathcal{S}_i$, $\mathcal{P}_i$, $\succ^i$)
	  \State $\mathcal{A}$ $\leftarrow$ $\mathcal{A}$ $\cup$ $\mathcal{A}_i$
        \EndFor
        \State return $\mathcal{A}$
  \State \textbf{end}
\end{algorithmic}
\end{algorithm}
In line 6, a doctor is randomly selected from the patient \emph{j}'s preference list and is held in $s^{*}$ data structure. Line 7 maintains the selected patient-doctor pairs of $x_i$ category in $\mathcal{A}_i$. Line 8 and 9 removes the selected patients and selected doctors from the system. Line 10 removes the selected doctor from the preference lists of the remaining patients. In line 11, the $p^*$ and $s^*$ are set to $\phi$. The \emph{RanPAM allocation} returns the final patient-doctor allocation pair set $\mathcal{A}_i$.  

\begin{algorithm}[H]
\caption{RanPAM allocation ($\mathcal{S}_i$, $\mathcal{P}_i$, $\succ^i$)}
\begin{algorithmic}[1]
      \Output $\mathcal{A}_i$ $\leftarrow$ $\phi$
	\State \textbf{begin}
	\State $j \leftarrow 0$, $p^{*}$ $\leftarrow$ $\phi$, $s^{*}$ $\leftarrow$ $\phi$
	      \While {$\mathcal{P}_i$ $\neq$ $\phi$} 
	      \State $j$ $\leftarrow$ $rand$($\mathcal{P}_i$)
	      \State $p^{*}$ $\leftarrow$ $p_{j}^{x_i}$
	      \State $s^*$ $\leftarrow$ $random$($\succ_j^i$) 
	      \State $\mathcal{A}_i$ $\leftarrow$ $\mathcal{A}_i$ $\cup$ $( p^{*}, s^{*})$
	      \State $\mathcal{P}_i$ $\leftarrow$ $\mathcal{P}_i$ $\setminus$  $p^{*}$
	      \State $\mathcal{S}_i$ $\leftarrow$ $\mathcal{S}_i$ $\setminus$  $s^{*}$
	      \State $\succ_k^{i}$ $\leftarrow$ $\succ_k^{i}$ $\setminus $ $s^{*}$, $\forall{k}$ $\in$ $\mathcal{P}_i$
	      \State $p^* \leftarrow \phi$, $s^* \leftarrow \phi$
	      \EndWhile
	      \State return $\mathcal{A}_i$
  \State \textbf{end}
\end{algorithmic}
\end{algorithm}
\noindent $-$ \textbf{Upper Bound Analysis:} \\
The time taken by the RanPAM is the sum of running times for each statement executed. Considering the \emph{Main()}, it can be seen that line 2-5 will execute for \emph{k} times. In \emph{RanPAM allocation}, line 2 is bounded by the constant time. In line 3, the test
is executed $(n + 1)$ times, as their are \emph{n} patients in $\mathcal{P}_i$ .
For each execution of while loop, line $4-9$  will take constant amount of time; whereas line 10 is bounded above by $n$. The overall time complexity of the RanPAM is mainly contributed by the \emph{while} loop in line 3-12.
Mathematically, the upper bound on the RanPAM for all the \emph{k} categories is given as:
\begin{equation*}
\begin{rcases}
 T(n) = \sum_{i=1}^{k}\Bigg(1 + \Bigg(\sum_{i=1}^{n} i-1 \Bigg)\Bigg)\\
 = \Bigg(\sum_{i=1}^{k} 1\Bigg) + \Bigg(\sum_{i=1}^{k}\sum_{i=1}^{n} i-1 \Bigg)\\
 \leq \Bigg(\sum_{i=1}^{k} 1\Bigg) + \Bigg(\sum_{i=1}^{k}\sum_{i=1}^{n} i\Bigg)\\
   = k  + \sum_{i=1}^{k} \frac{n(n+1)}{2}\\
= k  +  \frac{kn(n+1)}{2}\\
= \frac{kn^2 + k(n+2)}{2}\\
 T(n) = O\Bigg(kn^2\Bigg)
\end{rcases}
\end{equation*}
\noindent$-$ \textbf{Essential properties:}\\
There are two essential properties that will help to develop the further mechanisms in our paper. The two properties are: \emph{Blocking coalition} and \emph{Core allocation}. These properties captures the fundamental idea: can we design a system where agents cannot gain by manipulating their preferences that are only known to them?
\begin{itemize}
\item
\textbf{Blocking coalition}. For every $\mathcal{T}_i$ $\subset$ $\mathcal{P}_i$ let $\mathcal{A}_i(\mathcal{T}_i)$ = $\{u \in \mathcal{A}_i: u_i^{p_{i}^{x_i}} \in \mathcal{T}_i, ~\forall{i} \in \mathcal{T}_i\}$ denote the set of allocations that can be achieved by the agents in $\mathcal{T}_i$ trading among themselves alone. Given an allocation $a \in \mathcal{A}_i$, a set $\mathcal{T}_i \subseteq \mathcal{P}_i$ of agents is called a \emph{blocking coalition} (for \emph{a}), if there exists a $u \in \mathcal{A}(\mathcal{T})$ such that $\forall{i} \in \mathcal{T}$ either $u_i \succ_{i} a_i$ or $u_i = a_i$ and at least one agent is better off $i.e.$ for at least one $j \in \mathcal{T}_i$ we have $u_j \succ_j a_j$. 
\item
\textbf{Core allocation}. 
This property exhibits the fact that the allocation will be free of \emph{blocking coalition}. In other words it says that if any subset of agents form a coalition and reallocates themselves via some internal reallocation, all of the members of the coalition can't be better off.  
\end{itemize} 
\noindent $-$ \textbf{Drawback:}\\
 From the perspective of \emph{blocking coalition}, it can be concluded that the RanPAM is suffering from the blocking coalition. This leads to the violation of one of the economic properties in MDWM environment named as \emph{core allocation}.
\subsection{Truthful Optimal Allocation Mechanism (TOAM)}
The proposed truthful mechanism needs to overcome several non-trivial challenges: firstly, the patients preferences are unknown and need to be reported in a truthful manner; secondly, the allocation of doctors made to the patients must satisfy the \emph{core}. The previously discussed RanPAM mechanism failed to handle such challenges. To overcome these challenges, in this paper a \emph{truthful} mechanism is proposed which is termed as TOAM. Along with, \emph{truthfulness}, TOAM satisfies \emph{Pareto Optimality} (defined later). The main idea of the TOAM is to develop a mechanism where the agents can't gain by manipulation. If there is no manipulation we can reach to the equilibrium of the system very quickly and the market become stable. The TOAM satisfies two useful properties mentioned in the previous subsection. The two properties are:  
\begin{itemize}
\item
\textbf{Truthfulness or DSIC}. Let $\mathcal{A}_i$ = $\mathcal{M}(\succ_i^i, \succ_{-i}^i)$ and $\hat{\mathcal{A}_i} = \mathcal{M}(\hat{\succ}_{i}^i, \succ_{-i}^i)$. TOAM is \emph{truthful} if $a(i)^i \succeq_i^i \hat{a}(i)^i$, for all $p_{i}^{x_i} \in \mathcal{P}_i$.  


\item
\textbf{Pareto optimality}. An allocation $\mathcal{A}_i$ is \emph{pareto optimal} if there exists no allocation $a_{j}^i \in \mathcal{A}_i$ such that any patient $p_{i}^{x_i} \in a_{j}^{i}$ can make themselves better off without making other patient(s) $p_{k}^{x_i} \in a_{k}^{i}$ worse off.
\end{itemize}

\noindent $-$ \textbf{Sketch of the TOAM:}\\
More formally, the proposed TOAM, unlike the previous mechanism, can be thought of as a four stage allocation mechanism: \emph{Main routine}, \emph{Graph initialization}, \emph{Graph creation} and \emph{Cycle detection}.
\paragraph{\textbf{\underline{Main routine}}}
The main idea behind the construction of \emph{main routine} is to capture all the \emph{k} categories present in the system.
The input to the \emph{main routine} are the set of sets of vertices representing all the available patients given as $\mathcal{C} = \{\mathcal{C}_1, \mathcal{C}_2, \ldots, \mathcal{C}_k\}$; where $\mathcal{C}_i = \{c_1^{x_i}, c_{2}^{x_i}, \ldots, c_{n}^{x_i}\}$ is the set of vertices representing patients belonging to $x_i$ category, the set of sets of vertices representing all the available expert consultants (doctors) given as $\mathcal{Q} = \{\mathcal{Q}_1, \mathcal{Q}_2, \ldots, \mathcal{Q}_k\}$; where $\mathcal{Q}_i = \{q_{1}^{x_i}, q_{2}^{x_i}, \ldots, q_{n}^{x_i}\}$ is the set of vertices representing doctors belonging to $x_i$ category, and $x$ represents the set of categories. The output of the \emph{main routine} is the allocation set $\mathcal{A}$. In Line 4, the $\mathcal{C}^*$ data structure temporarily holds the set of vertices returned by \emph{select()} in $x_i$ category. The $\mathcal{Q}^*$ data structure temporarily holds the set of vertices returned by \emph{select()} in $x_i$ category as depicted in line 5. In line 6 for each category $x_i$, a call to \emph{graph initialization} is made to randomly allocate a doctor to each patient. The allocation set $\mathcal{A}$ maintains the allocation for each category in line 7. In line 9, the final allocation set $\mathcal{A}$ is returned.
   \begin{algorithm}[H]
\caption{Main routine ($\mathcal{C}$, $\mathcal{Q}$, $x$, $\succ$)}
\begin{algorithmic}[1]
\Output $\mathcal{A}$ $\leftarrow$ $\phi$
\State \textbf{begin}
\State $\mathcal{Q}^*$ $\leftarrow$ $\phi$, $\mathcal{C}^*$ $\leftarrow$ $\phi$
\For{each $x_i \in x$}
   \State $\mathcal{C}^*$ $\leftarrow$ \emph{select}($\mathcal{C}$)
   \State $\mathcal{Q}^{*}$ $\leftarrow$ \emph{select}($\mathcal{Q}$)
   \State $\mathcal{A}_i$ $\leftarrow$ Graph initialization ($\mathcal{C}^*$, $\mathcal{Q}^*$, $\succ^i$)
   \State $\mathcal{A}$ $\leftarrow$ $\mathcal{A}$ $\cup$ $\mathcal{A}_i$
\EndFor   
\Return $\mathcal{A}$
\State \textbf{end}
\end{algorithmic}
\end{algorithm}       

\paragraph{\underline{\textbf{Graph initialization}}}
The input to the \emph{graph initialization} phase are the set of vertices representing the patients in $x_i$ category $i.e.$ $\mathcal{C}_i$, the set of vertices representing the doctors in $x_i$ category $i.e.$ $\mathcal{Q}_i$, and the preference profile of the patients in $x_i$ category. 
The output of the \emph{graph initialization}
is the graph $\mathcal{G}$ in the form of adjacency matrix $\mathcal{F}$ representing the randomly allocated doctors to the patients. Line 2, initializes the adjacency matrix $\mathcal{F}$ of size $|\mathcal{V}_i| * |\mathcal{V}_i|$ to null matrix; where $\mathcal{V}_i = \mathcal{C}_i \cup \mathcal{Q}_i$.
\begin{algorithm}[H] 
\caption{Graph initialization ($\mathcal{C}_i$, $\mathcal{Q}_i$, $\succ^i$)}
\begin{algorithmic}[1]
 	      \State \textbf{begin}
 	      \State $\mathcal{F}$ = $\{0\}_{|\mathcal{V}_i|*|\mathcal{V}_i|}$ 
 	      \For {each vertex $c_{t}^{x_i}$ $\in \mathcal{C}_i$} 
 	           \State $q^*$ $\leftarrow$ $Select\_random(\mathcal{Q}_i)$
 	           \State $\mathcal{F}_{q^{*}, c_t^{x_i}}$ = 1  
 	           \State $\mathcal{Q}_i$ $\leftarrow$ $\mathcal{Q}_i \setminus q^{*}$ 
 	      \EndFor
 	      \State Graph creation ($\mathcal{C}_i$, $\mathcal{Q}_i$, $\mathcal{F}$, $\succ^i$)  
             \State \textbf{end}
\end{algorithmic}
\end{algorithm}
The \emph{for} loop in line 3 iterates over all the patients in the $x_i$ category. 
In line 4, the \emph{Select\_random()} function takes the set of vertices $\mathcal{Q}_i$ (analogous to the doctors with $x_i$ expertise area) as the input and returns the randomly selected vertex. The randomly selected vertex is held in $q^*$ data structure.
Line 5, places a directed edge from $q^{*}$ to $c_t^{x_i}$. Line 6, removes the randomly allocated vertex held in $q^*$ from $\mathcal{Q}_i$. In line 8, a call to \emph{Graph creation} phase is done.
\paragraph{\underline{\textbf{Graph creation}}} 
The input to the \emph{graph creation} phase are the set of vertices representing the patients in $x_i$ category $i.e.$ $\mathcal{C}_i$, the set of vertices representing the doctors in $x_i$ category $i.e.$ $\mathcal{Q}_i$, the adjacency matrix $\mathcal{F}$, and the preference profile of the patients in $x_i$ category. 
The output of the \emph{graph creation} is the adjacency matrix $\mathcal{F}$. In line 3, the \emph{Select\_best()} function takes the strict preference ordering list of $t^{th}$ agent as input and returns the best doctor from the available preference list. The $q^{*}$ data structure holds the best selected doctor. Line 4 places a directed edge from $c_t^{x_i} \in \mathcal{C}_i$ to $q^* \in Q_i$. In line 6, a call to \emph{Optimal allocation} phase is done. 
\begin{algorithm}[H] 
 \caption{Graph creation ($\mathcal{C}_i$, $\mathcal{Q}_i$, $\mathcal{F}$, $\succ^i$)}
 \begin{algorithmic}[1]
 	      \State \textbf{begin}
 	      \For {each vertex $c_{t}^{x_i} \in \mathcal{C}_i$}
                      \State $q^*$ $\leftarrow$ $Select\_best(\succ_t^i)$
                      \State $\mathcal{F}_{c_{t}^{x_i}, q^{*}}$ = 1
              \EndFor
              \State Optimal allocation ($\mathcal{C}_i$, $\mathcal{Q}_i$, $\mathcal{F}$)   
              \State \textbf{end}
 \end{algorithmic}
 \end{algorithm} 
\paragraph{\underline{\textbf{Optimal allocation}}}
The next challenge is to determine a finite cycle in a directed graph $\mathcal{G}$. The input to the \emph{optimal allocation} phase are the set of vertices representing the patients in $x_i$ category $i.e.$ $\mathcal{C}_i$, the set of vertices representing the doctors in $x_i$ category $i.e.$ $\mathcal{Q}_i$, and the adjacency matrix $\mathcal{F}$ returned from the previous stage. Initially, in line 3-5 all $c_k^{x_i} \in \mathcal{C}_i$ and $q_{j}^{x_i} \in \mathcal{Q}_i$ are marked unvisited. In line 6 a random vertex from set $\mathcal{V}_i$ is selected and is captured by $\pi$ data structure. Line 7 marks the vertex in $\pi$ as visited. The vertex in $\pi$ is pushed into the stack $S$ using line 8. Now, Line $9-25$, computes a finite directed cycle in the graph by following the outgoing arcs. Line $26$, reallocates as suggested by directed cycle. Each patient on a directed cycle gets the expert consultant better than the expert consultant it initially points to or the initially pointed expert consultant. Line 27-38 removes the patients and doctors that were reallocated in the previous step from the available patients and doctors lists respectively and updates the adjacency matrix. Line 42-50 updates the available patient and doctors list, preference list of the patients, and the adjacency matrix. A call is made to the \emph{graph creation} phase to generate the updated graph from the available number of patients and the expert consultants until the patients set and doctor sets are not empty.\\

\noindent $-$ \textbf{Several properties of TOAM:}\\
The proposed TOAM has several compelling properties. These properties are discussed next.\\
\\
$\bullet$ \textbf{Running time}
The running time of TOAM will be the sum of the running time of \emph{main routine}, \emph{graph initialization}, \emph{graph creation}, and \emph{optimal allocation} phases. Line 2 of the \emph{main routine} is bounded by $O(1)$. The \emph{for} loop in line 3 executes for $k + 1$ times.
Line 4-5 are bounded by $O(1)$ for each iteration of the \emph{for} loop. Line 6, takes the time equal to the time taken by \emph{graph 
initialize} mechanism. For the time being, let the graph initialization mechanism is bounded by $O(N)$. Line 7 of \emph{main routine} mechanism takes $O(1)$ time. So, the running time of \emph{main
 routine} is bounded by: $O(1) + O(kN) + O(1) = O(kN)$. In \emph{graph initialization},  in each execution of the \emph{for} loop an edge is placed between the two vertices of the graph $\mathcal{G}
 $. 
\begin{algorithm}[H] 
 \caption{Optimal allocation ($\mathcal{C}_i$, $\mathcal{Q}_i$, $\mathcal{F}$)}
 \begin{algorithmic}[1]
 	      \State \textbf{begin} 
 	      \State $S$ $\leftarrow$ $\phi$, $\pi \leftarrow \phi$, $\hat{\mathcal{C}}^* \leftarrow \phi$, $\hat{\mathcal{Q}}^* \leftarrow \phi$
             \For {each $c_{k}^{x_i}$ $\in$ $\mathcal{C}_i$ and $q_{j}^{x_i}$ $\in$ $\mathcal{Q}_{i}$}
                 \State Mark $c_{k}^{x_i}$ $\leftarrow$ unvisited, $q_{j}^{x_i}$ $\leftarrow$ unvisited 
             \EndFor
              \State $\pi$ $\leftarrow$ $random(\mathcal{V}_i)$
              \State Mark $\pi$ $\leftarrow$ visited
              \State $push (S, \pi)$
              \While {S is non-empty}
                  \State $\pi \leftarrow pop(S)$
                    \For {each $\pi'$ adjacent to $\pi$}
                           \If {$\pi'$ is unvisited}
                                \State Mark $\pi'$ $\leftarrow$ visited
                                \State $push (S, \pi')$
                           \ElsIf {$\pi'$ is visited}
                               \State Exists a finite cycle.
                               \For{each $c_{k}^{x_i}$ in finite cycle}
                                  \State $\hat{\mathcal{C}}^*$ $\leftarrow$ $\hat{\mathcal{C}}^*$ $\cup$ $\{c_{k}^{x_i}\}$
                               \EndFor
                               \For{each $q_{j}^{x_i}$ in finite cycle}
                                  \State $\hat{\mathcal{Q}}^*$ $\leftarrow$ $\hat{\mathcal{Q}}^*$ $\cup$ $\{q_{j}^{x_i}\}$
                               \EndFor   
                           \EndIf
                    \EndFor
             \EndWhile 
             \State Allocate each patient in cycle a doctor it points to.
            \State  $\mathcal{C}_i$ $\leftarrow$ $\mathcal{C}_i \setminus \hat{\mathcal{C}}^*$
             \State $\mathcal{Q}_i \leftarrow \mathcal{Q}_i \setminus \hat{\mathcal{Q}}^*$
             \For{each $c_{k}^{x_i}$ $\in$ $\hat{\mathcal{C}}^*$}
              \For{each $q_j^{x_i}$ $\in$ $\hat{\mathcal{Q}}^*$}
              \If{ $\mathcal{F}_{c_{k}^{x_i}, q_j^{x_i}} == 1$}
                              \State $\mathcal{F}_{c_{k}^{x_i},q_j^{x_i}} = 0$
                        \EndIf
              \If{ $\mathcal{F}_{q_{j}^{x_i}, c_k^{x_i}} == 1$}
                              \State $\mathcal{F}_{q_{j}^{x_i},c_k^{x_i}} = 0$
                        \EndIf          
              \EndFor
             \EndFor
            \State $\mathcal{V}_i = \mathcal{Q}_i \cup \mathcal{C}_i$  \Comment{$\mathcal{V}_i$ is updated}
             \If {$\mathcal{V}_i \neq \phi$}
                \State $\succ^i$ $\leftarrow$ $\phi$
                \For {all $c_{k}^{x_i} \in \mathcal{C}_i$}
                    \State $\succ_{k}^{i} \leftarrow \succ_{k}^i \setminus \hat{\mathcal{Q}}^*$ \Comment{Removes doctors present in $\hat{\mathcal{Q}}^*$ from \hspace*{33mm} patient's $k$ preference list.}
                    \State $\succ^i$ $\leftarrow$ $\succ^i$ $\cup$ $\succ_{k}^{i}	$
                    \For {all $q_j^{x_i} \in \hat{\mathcal{Q}}^*$}
                        \If{ $\mathcal{F}_{c_{k}^{x_i}, q_j^{x_i}} == 1$}
                              \State $\mathcal{F}_{c_{k}^{x_i},q_j^{x_i}} = 0$
                        \EndIf
                    \EndFor
                \EndFor             
                \State Graph creation ($\mathcal{C}_i$, $\mathcal{Q}_i$, $\mathcal{F}$, $\succ^i$)  
             \EndIf
       \State \textbf{end}
 \end{algorithmic}
 \end{algorithm} 
Generating a directed graph $\mathcal{G}$ using line $3-7$ takes $O(n)$ time. Next, line 10 is bounded by the time taken by the \emph{graph creation} mechanism. In the \emph{graph creation} algorithm, the \emph{for} loop contributes the major part of the running time $i.e.$ $O(n)$. Line 6 of \emph{graph creation} is bounded by the time taken by \emph{optimal allocation}. For the time being let the time taken by by the \emph{optimal allocation} be $O(M)$. So, the running time of 
 graph creation algorithm is bounded by: $O(1) + O(n) + O(M)$. In, optimal allocation algorithm line $2$ is bounded by $O(1)$. The total number of vertex is $n + n = 2n$, so the outer \emph{for} 
 loop will take $O(n)$. Line $6-8$, is bounded by $O(1)$. the total number iterations of the innermost \emph{while} loop of optimal allocation cannot exceed the number of edges in $\mathcal{G}$, and 
 thus the size of \emph{S} cannot exceed \emph{n}. The \emph{while} loop of optimal allocation algorithm is bounded by $O(n)$. Line $26-28$, are executed in constant time $O(1)$. Line $29-38$ of the 
 mechanism can be executed in worst case $O(n^2)$. Line 42-50 in worst case is bounded by $O(n)$.
 The running time of \emph{optimal allocation} is: $O(n) + O(1) + O(n^2) + O(1) + O(n) = O(n^2)$. 
 Total running time of TOAM: $O(n) + O(n^2) = O(n^2)$. Considering the \emph{k} categories simultaneously we have $O(kn^2)$.
 \\
 \\
 $\bullet$ \textbf{DSIC}
The second property, that distinguishes the proposed TOAM from any direct revelation allocation mechanism is its DSIC property. In TOAM, the strict preference ordering revealed by the agents in any category $x_i \in x$ over the set of doctors $\mathcal{S}_i$ are unknown or private to the agents. As the strict preference ordering is private, any agent \emph{i} belonging to category $x_j \in x$ can misreport their private information to make themselves better off. TOAM, an obvious direct revelation mechanism claims that agents in any category $i \in 1 \dots k$ cannot make themselves better off by misreporting their private valuation, i.e. TOAM is DSIC. 
\newtheorem*{Theo1}{Theorem 1}
\begin{Theo1}
The TOAM is DSIC.
\end{Theo1}
\begin{proof}
The truthfulness of the TOAM is based on the fact that each agent \emph{i} gets the best possible choice from the reported strict preference, irrespective of the category $i \in 1 \dots k$  of the agent \emph{i}. It is to be noted that, the third party (or the platform) partition the available patients and doctors into different sets based on their category. The partitioning of doctors set $S =\{\mathcal{S}_1, \mathcal{S}_2, \dots, \mathcal{S}_k\}$ is independent of the partitioning of the available patients into the set $\mathcal{P} = \{\mathcal{P}_1, \mathcal{P}_2, \dots, \mathcal{P}_k\}$. So, if we select the patient set $\mathcal{P}_i \in \mathcal{P}$ and the doctor set $\mathcal{S}_i \in \mathcal{S}$ randomly from category $x_i \in x$ and show that for any agent $p_{i}^{x_i} \in \mathcal{P}_i$ misreporting the private information (in this case strict preference over $\mathcal{S}_i$) will not make the agent $p_{i}^{x_i}$ better-off, then its done. Our claim is that, if any agent belonging to $x_i$ category, cannot be better off by misreporting their strict preference, then no agent from any category can be better off by misreporting the strict preference.\\ 
\indent Fix category $x_i$. Let us assume that, if all the agents in $x_i$ are reporting \emph{truthfully}, then all the agents gets a doctor till the end of $m^{th}$ iteration. From the construction of the mechanism in each iteration of the TOAM, at least a cycle $\Omega_i \in \Omega$ is selected. The set of cycles chosen by the TOAM in $m$ iterations are: $\Omega = (\Omega_1, \Omega_2, \ldots, \Omega_m)$, where $\Omega_i$ is the cycle chosen by the TOAM in the $i^{th}$ iteration, when agents reporting \emph{truthfully}. Each agent in $\Omega_1$, gets its first choice and hence no strategic agent can be benefited by misreporting. From the construction of the mechanism, no agent in $\Omega_i$ will ever be pointed by any agents in $\Omega_1, \ldots, \Omega_{i-1}$; if this is not the case, then agent \emph{i} could have been belong to one of the previously selected cycle.\\
Once the doctor is allocated to the agent, the mechanism remove the agent along with the allocated doctor, and the strict preference list of the remaining agents are updated. Since, agent \emph{i} gets its first choice outside of the doctors allocated in $\Omega_1, \ldots, \Omega_{i-1}$, it has no incentive to \emph{misreport}. Thus, whatever agent \emph{i} reports, agent \emph{i} will not receive a doctor owned by an agent in $\Omega_1, \ldots, \Omega_{i-1}$. Since, the TOAM gives agent \emph{i} its favourite doctor outside the selected cycle till now. 
 Hence, agent \emph{i} did not gain by misreporting the strict preference ordering. From our claim it must be true for any agents in any category $i \in 1 \dots k$. Hence, TOAM is DSIC. 
      
\end{proof}

$\bullet$ \textbf{Core allocation}
The third property exhibited by the proposed TOAM is related to the \emph{uniqueness} of the resultant allocation or in some sense \emph{optimality}. The term used to determine the optimal allocation of TOAM is termed as \emph{unique core allocation}. The claim is that, the allocation computed by the proposed TOAM is the \emph{unique core allocation}. 
\newtheorem*{Theo2}{Theorem 2} 
\begin{Theo2}
The allocation computed by TOAM is the \emph{unique} \emph{core} \emph{allocation}.

\end{Theo2}
\begin{proof}
The proof of \emph{unique core allocation} for any category $x_i \in x$ can be thought of as divided into two parts. First, it is proved that the allocation vector $\mathcal{A}_i$ computed by TOAM for any category $x_i \in x$ is a \emph{core allocation}. Once the allocation vector in $x_i$ category is proved to be \emph{The core}, the \emph{uniqueness} of the \emph{core} allocation for $x_i$ is taken into consideration. Our claim is that, if the allocation $\mathcal{A}_i$ computed by TOAM for any arbitrary $x_i \in x$ is a \emph{unique core allocation}, then the allocation computed by TOAM for all $x_i \in x$ will be a unique core allocation. \\
\indent Fix category $x_i$. In order to prove the allocation computed by TOAM is a \emph{core} allocation, consider an arbitrary sets of agents $S^*$, such that $S^* \subseteq P$. Let $\Omega_i$ is the cycle chosen by TOAM in the $i^{th}$ iteration and $\delta(\Omega_i)$ is the set of agents allocated a doctor, when reporting \emph{truthfully}. When TOAM will allocate the agents, at some cycle $\Omega_k$, $i \in S^*$ will be included for the first time. In that case $\delta(\Omega_k) \cap S^* \neq \phi$. As any agent $i \in S^*$ is being included for the first time, it can be said that no other agent in $S^*$ is included in the cycles $\Omega_1, \dots, \Omega_{k-1}$. As the TOAM allocates the favourite doctor to any arbitrary agent $i \in \delta(\Omega_k)$ outside the doctors allocated to $\delta(\Omega_1), \dots, \delta(\Omega_{k-1})$, it can be concluded that $i \in \delta(\Omega_k)$ and $i \in S^*$ such that $\delta(\Omega_k) \cap S^* \neq \phi$ gets his favourite doctor at the $k^{th}$ iteration. Hence no internal reallocation can provide a better doctor to any agent $i \in S^*$. Inductively, the same is true for any agent $j \in S^*$ that will satisfy $\delta(\Omega_k) \cap S^* \neq \phi$.\\
Now, we prove \emph{uniqueness}. In TOAM, each agent in $\Omega_1$ receives the best possible doctor from his preference list. Any core allocation must also do the same thing, otherwise the agents who didn't get the first choice could be better off with internal reallocation. So the core allocation agrees with the TOAM allocation for the agents in $\delta(\Omega_1)$. Now in TOAM, as all the agents in $\delta(\Omega_2)$ get their favourite doctors outside the set of doctors allocated to the agents $\delta (\Omega_1)$, any core allocation must be doing the same allocation, otherwise the agents in $\delta(\Omega_2)$ who didn't get their favourite choice can internally reallocate themselves in a better way. In this way we can inductively conclude that the core allocation must follow the TOAM allocation. This proves the uniqueness of TOAM. \\
\indent Hence, it is proved that the allocation by TOAM for category $x_i$ is a \emph{unique} \emph{core} \emph{allocation}. From our claim it must be true for any agents in categories $i \dots k$. Hence,the allocation computed by TOAM for any category $x_i \in x$ is the \emph{unique core allocation}.      
\end{proof}
\noindent $-$ \textbf{Correctness of the TOAM:}\\
The correctness of the TOAM mechanism is proved with the \emph{loop invariant} technique \cite{Coreman_2009} \cite{Gries_2009}. The \emph{loop invariant}: At the start of $j^{th}$ iteration, the number of patient-doctor pairs to be explored are $n - \sum_{i=1}^{j-1} k_i$ in a category, where $k_i$ is the number of patient-doctor pairs processed at the $i^{th}$ iteration. Precisely, it is to be noted that $n - \sum_{i=1}^{j-1} k_i \leq n$. From definition of $k_i$, it is clear that the term $k_i$ is non-negative. The number of patient-doctor pairs could be atleast 0. Hence, satisfying the inequality $n - \sum_{i=1}^{j-1} k_i \leq n$. We must show three things for this loop invariant to be true.
\paragraph*{\textbf{Initialization}}
It is true prior to the first iteration of the loop. Just before the first iteration of the \emph{while} loop, in optimal allocation mechanism $n - \sum_{i=1}^{j-1} k_i \leq n$ $\Rightarrow$ $n - 0 \leq n$ $i.e.$ no patient-doctor pair is explored apriori in, say $i^{th}$ category. This confirms that $\mathcal{A}_i$ contains no patient-doctor pair.
\paragraph*{\textbf{Maintenance}}
For the loop invariant to be true, if it is true before each iteration of \emph{while} loop, it remains true before the next iteration. The body of \emph{while} loop allocates doctor(s) to the patient(s) with each doctor is allocated to one patient present in the detected cycle; $i.e.$ each time $\mathcal{A}_i$ is incremented or each time \emph{n} is decremented by $k_i$. Just before the $j^{th}$ iteration the number of patient-doctor pairs allocated are $\sum_{i=1}^{j-1} k_i$, implies that the number of patient-doctor pairs left are: $n - \sum_{i=1}^{j-1}k_i \leq n$. After the $j^{th}$ iteration, two cases may arise:\\
\textbf{Case 1:} If $k_j = n - \sum_{i=1}^{j-1} k_i$\\
\indent In this case, all the $k_j$ patient-doctor pairs will be exhausted in the $j^{th}$ iteration and no patient-doctor pair is left for the next iteration. The inequality $n - (\sum_{i=1}^{j-1} k_i + k_j)$ = $ (n - \sum_{i=1}^{j-1} k_i) - k_j$ = ($n - \sum_{i=1}^{j-1} k_i$) $-$ ($n - \sum_{i=1}^{j-1} k_i$) = 0 $\leq$ $n$.\\
\textbf{Case 2:}
If $k_j < n - \sum_{i=1}^{j-1} k_i$\\
\indent In this case, $j^{th}$ iteration allocates few patient-doctor pairs
 from the remaining patient-doctor pairs; leaving behind some of the pairs for further iterations. So, the inequality $n - (\sum_{i=1}^{j-1}k_i + k_j) \leq n$ = $n - \sum_{i=1}^{j}k_i \leq n$ is satisfied.\\
\indent From Case 1 and Case 2, at the end of $j^{th}$ iteration the loop invariant is satisfied.
\paragraph*{\textbf{Termination}}
In each iteration at least one patient-doctor pair is formed. This indicates that at some $(j+1)^{th}$ iteration the loop terminates and in line no. 9, $S$ is exhausted, otherwise the loop would have continued. As the loop terminates and $S$ is exhausted in $(j+1)^{th}$ iteration. We can say $n - \sum_{i=1}^{j} k_i = 0 \leq n$. Thus indicates that all the \emph{n} agents are processed and each one has a doctor assigned when the loop terminates. \\
\\ 
\noindent $-$ \textbf{Illustrative example:}\\
The number of patients is $n =5$ $i.e.$ $\mathcal{P}_2 = \{p_1^{x_2}, p_2^{x_2}, p_3^{x_2}, p_4^{x_2}, p_{5}^{x_2}\}$ and the number of expert consultant (or doctors) is $n=5$ $\mathcal{S}_2 = \{s_1^{x_2}, s_2^{x_2}, s_3^{x_2}, s_4^{x_2}, s_{5}^{x_2}\}$. The strict preference ordering given by the patient set $\mathcal{P}_2$ is: $p_1^{x_2}$: ($s_2^{x_2}$, $s_4^{x_2}$, $s_3^{x_2}$, $s_1^{x_2}$, $s_{5}^{x_2}$); $p_2^{x_2}$: ($s_3^{x_2}$, $s_4^{x_2}$, $s_5^{x_2}$, $s_1^{x_2}$, $s_{2}^{x_2}$); $p_3^{x_2}$: ($s_2^{x_2}$, $s_3^{x_2}$, $s_1^{x_2}$, $s_4^{x_2}$, $s_{5}^{x_2}$); $p_4^{x_2}$: ($s_5^{x_2}$, $s_2^{x_2}$, $s_3^{x_2}$, $s_4^{x_2}$, $s_{1}^{x_2}$); $p_5^{x_2}$: ($s_1^{x_2}$, $s_4^{x_2}$, $s_2^{x_2}$, $s_3^{x_2}$, $s_{5}^{x_2}$). Following the \emph{graph initialization} phase a directed edge is placed between the following pairs:  $\{(s_1^{x_2}, p_{1}^{x_2}), (s_2^{x_2}, p_{2}^{x_2}), (s_3^{x_2}, p_{3}^{x_2}), (s_4^{x_2}, p_{4}^{x_2}), and (s_5^{x_2}, p_{5}^{x_2})\}$. Now, Following the \emph{graph creation} phase, say, a patient $p_1^{x_2}$ is selected from $\mathcal{P}_2$. As, $s_2^{x_2}$ is the most preferred doctor in the preference list of $p_1^{x_2}$. So, a directed edge is placed from $p_{1}^{x_2}$ to $s_2^{x_2}$. The \emph{for} loop of the \emph{graph creation} phase places a directed edge between the remaining patients in $\mathcal{P}_2$ and the most prefered doctors in $\mathcal{S}_2$, resulting in a directed graph. Now, running the optimal allocation phase on the directed graph a cycle $(p_{2}^{x_2}, s_3^{x_2}, p_3^{x_2}, s_2^{x_2}, p_2^{x_2})$ is determined. Similarly, the remaining patients $\mathcal{P}_2 = \{p_1^{x_2}, p_4^{x_2}, p_5^{x_2}\}$ will be allocated a doctor. The final allocation of $patient-doctor$ pair are: $\{(p_1^{x_2}, s_4^{x_2}), (p_2^{x_2}, s_3^{x_2}), (p_3^{x_2}, s_2^{x_2}), (p_4^{x_2}, s_5^{x_2}), (p_5^{x_2}, s_1^{x_2})\}$.

\section{More general setting}
Till now, for simplicity in any category $x_i$ we have considered the set-up where the number of patients and number of doctors are same $i.e.$ \emph{n} along with an extra constraint that the patients are providing the strict preference ordering over all the available expert consultants. But, one can think of the situation where there are \emph{n} number of patients and \emph{m} number of doctors in any given category such that $m \neq n$ ($m >n$ or $m<n$). Moreover, the constraint that each of the patient is providing the strict preference ordering over all the available doctors is not essential and can be relaxed for all the three different set-ups ($i.e.$ $m==n$, $m<n$, and $m>n$). By relaxation, we mean that the subset of the available patients may give the strict preference ordering over the subset of the available doctors.   
\subsection{Truthful Optimal Allocation Mechanism for InComplete Preference (TOAM-IComP)}
As an extension of TOAM the TOAM-IComP is proposed motivated by \cite{T.roughgarden_2016}\cite{Manlove_2016} to cater the need of more realistic incomplete (or partial) preferences. It is to be noted that, along with \emph{truthfulness} the TOAM-IComP satisfies the previously discussed two economic properties: \emph{pareto optimality} and \emph{the core}. \\

\noindent $-$ \textbf{Sketch of the TOAM-IComP}\\ 
The input to the TOAM-IComP are: the set of \emph{n} available patients in a particular category $x_i$, the set of \emph{m} available doctors in a particular category $x_i$, and the set of preferences of all the patients for the available doctors in a $x_i$ category. The output of the TOAM-IComP is the allocated patient-doctor pairs. 
In line 2, all the variables and data structures are initialized to 0 and $\phi$ respectively. In line 3-5 numbers 1 to \emph{n} are captured in $\mathcal{B}$ data structure. Next, the generated list $\mathcal{B}$ is randomized using line 6-8. Line 9-12 assigns the distinct random numbers between 1 and \emph{n} stored in $\mathcal{B}$ to the patients sequentially. In line 13, the patient list $\mathcal{P}_i$ is sorted based on the assigned random numbers. From line 14, it is clear that the mechanism terminates, once the patient list becomes empty. In line 15, using \emph{pick()} function, patient is selected sequentially based on the number assigned. Line 16 checks the preference list of patient stored in $\hat{p}$. In line 17, the best available doctor is selected from the selected patient preference list by using $Select\_best()$ function. Line 18 maintains the selected patient-doctor pairs in $\mathcal{F}$ data structure. Line 19 and 20 removes the selected patients and selected doctors from their respective preference lists. Line 22 sets $\hat{p}$ and $\hat{s}$ to $\phi$. The TOAM-IComP returns the final patient-doctor pair allocation set $\mathcal{F}$.
 \begin{algorithm}[H]
\caption{TOAM-IComP ($\mathcal{S}_i$, $\mathcal{P}_i$, $\succ^i$)}
\begin{algorithmic}[1]
      \Output $\mathcal{F}$ $\leftarrow$ $\phi$.
	\State \textbf{begin}
	\State $\ell$ $\leftarrow$ $0$, $\hat{p}$ $\leftarrow$ $\phi$, $\hat{s}$ $\leftarrow$ $\phi$, $\mathcal{B} \leftarrow \phi$
	\For{$i$ = 1 to $n$}
	\State $\mathcal{B} \leftarrow \mathcal{B} \cup \{i\}$
	\EndFor
        \For{$i$ = 1 to $n$}
	\State swap $\mathcal{B}[i]$ with $\mathcal{B}[Random(i,n)]$
	\EndFor	
	\For{each $p_{j}^{x_i} \in \mathcal{P}_i$}
	\State $Assign(p_j^{x_i}, \mathcal{B}[\ell])$ 
	\State $\ell \leftarrow \ell+1$
	\EndFor
	\State $\mathcal{P}_i$ $\leftarrow$ Sort($\mathcal{P}_i$, $\mathcal{B}$) \Comment{Sort $\mathcal{P}_i$ based on random number \hspace*{37mm} generated.}
	\While {$\mathcal{P}_i$ $\neq$ $\phi$} 
	      \State $\hat{p}$ $\leftarrow$ $pick$($\mathcal{P}_i$) \Comment{Sequentially picks the patients based on \hspace*{31mm}the random number assigned.} 
	      \If{$\succ_j^{i}$ $\neq$ $\phi$} \Comment{where, $j=1,~ 2,\ldots, n$}
	      \State $\hat{s}$ $\leftarrow$ $Select\_best$($\succ_j^{i}$) 
	      \State $\mathcal{F}$ $\leftarrow$ $\mathcal{F}$ $\cup$ $( \hat{p}, \hat{s})$
	      \State $\mathcal{P}_i$ $\leftarrow$ $\mathcal{P}_i$ $\setminus$  $\hat{p}$
	      \State $\mathcal{S}_i$ $\leftarrow$ $\mathcal{S}_i$ $\setminus$  $\hat{s}$
	      \EndIf
	      \State $\hat{p}$ $\leftarrow$ $\phi$, $\hat{s}$ $\leftarrow$ $\phi$
 	 \EndWhile
	      \Return $\mathcal{F}$
  \State \textbf{end}
\end{algorithmic}
\end{algorithm}

\begin{table*}
\centering
\caption{Running time and Economic properties of the proposed mechanisms}
\label{my-label}
\begin{tabular}{|l|l|l|l|l|}
\hline
\multicolumn{2}{|l|}{\cellcolor[HTML]{FFFFFF}} & \multicolumn{3}{l|}{\hspace*{18mm}\textbf{Economic properties}} \\ \hline
          \textbf{Proposed mechanisms}             & \textbf{Running time}                       &  \textbf{The Core}       & \textbf{Truthfulness}       & \textbf{Pareto optimality}       \\ \hline
               RanPAM        & \hspace*{6mm} $O(kn^2)$                     & \hspace*{6mm} \xmark       & \hspace*{6mm} \xmark    &\hspace*{10mm}  \xmark     \\ \hline
               TOAM        &  \hspace*{6mm} $O(kn^2)$                    & \hspace*{6mm} \cmark      & \hspace*{6mm}  \cmark     & \hspace*{10mm} \cmark     \\ \hline
               TOAM-IComP &   \hspace*{6mm}   $O(kn^2)$                     & \hspace*{6mm} \cmark &  \hspace*{6mm} \cmark    &   \hspace*{10mm} \cmark         \\ \hline
\end{tabular}
\end{table*} 
\noindent $-$ \textbf{Upper Bound Analysis}\\
The \emph{random} number generator in line 3-12 is motivated by \cite{Coreman_2009} and is bounded above by \emph{n}. When a \emph{while} loop exits in the usual way (i.e., due to the inner loop header), the test is executed one time more than the body of the \emph{while} loop. 
\begin{equation*}
\small
\begin{rcases}
 T(n) =\sum_{i=1}^{k}\Bigg( \Bigg(\sum_{i=1}^{n} O(1) \Bigg) + \Bigg(O(n\lg n)\Bigg) + \Bigg(\sum_{i=1}^{n} O(n)\Bigg)\Bigg)\\
 =\Bigg(\sum_{i=1}^{k} \sum_{i=1}^{n} O(1) \Bigg) + \Bigg(\sum_{i=1}^{k}  O(n\lg n) \Bigg) + \Bigg(\sum_{i=1}^{k} \sum_{i=1}^{n} O(n) \Bigg) \\
= O\Bigg(\sum_{i=1}^{k} \sum_{i=1}^{n} 1 \Bigg) + O\Bigg(\sum_{i=1}^{k}  n\lg n \Bigg) + O\Bigg(\sum_{i=1}^{k} \sum_{i=1}^{n} n \Bigg) \\
= O\Bigg(\sum_{i=1}^{k} n \Bigg) + O\Bigg(  kn\lg n \Bigg) + O\Bigg(\sum_{i=1}^{k}  n^2 \Bigg) \\
= O\Bigg(kn \Bigg) + O\Bigg(  kn\lg n \Bigg) + O\Bigg(kn^2\Bigg) \\
 T(n) = O\Bigg(kn^2\Bigg) 
\end{rcases}
\end{equation*}
In line 14, the test is executed $(n+1)$ times, as their are \emph{n} patients in $\mathcal{P}_i$. In line 13, the sorting is done that is bounded above by $n\lg n$. 
For each execution of \emph{while} loop line $14-23$ will take constant amount of time.\\

\begin{figure}[H]
        \begin{subfigure}[b]{0.29\textwidth}
                \includegraphics[scale=0.6]{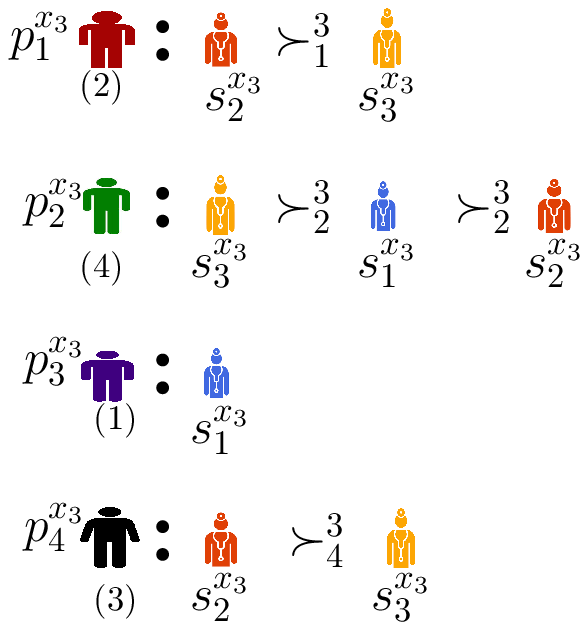}
                \caption{Strict preference ordering}
                \label{fig:gull}
        \end{subfigure}%
        \begin{subfigure}[b]{0.20\textwidth}
                \includegraphics[scale=0.6]{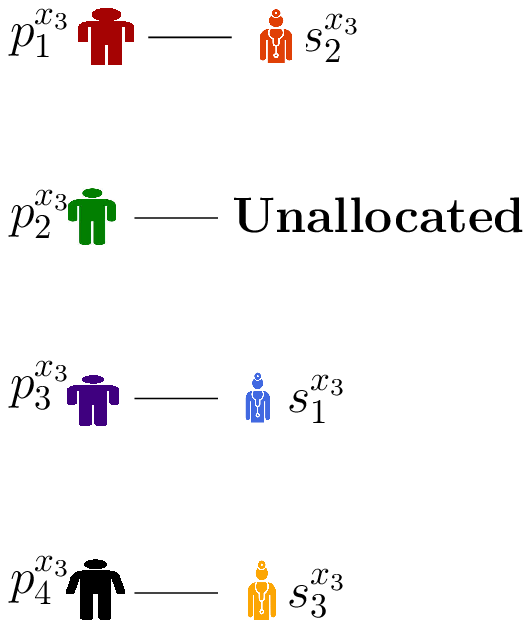}
                \caption{Final allocation}
                \label{fig:gull2}
        \end{subfigure}
        \caption{Detailed functioning of TOAM-IComP}\label{fig:animals}
\end{figure}

\noindent $-$ \textbf{Illustrative example}\\
The detailed functioning of TOAM-IComP for category $x_3$ is illustrated in Fig. \ref{fig:animals}. The number of patients is $n =4$ and the number of expert consultants (or doctors) is $m=3$. 
The strict preference ordering given by the patient set $\mathcal{P}_3$ is shown in Fig. \ref{fig:gull}. Following line 3-12 of Algorithm 6, we generate the random numbers for the patients in $\mathcal{P}_3$. Now, based on the random number assigned as shown in Fig. \ref{fig:gull}, first the patient $p_{3}^{x_3}$ is selected and assigned expert consultant $s_{3}^{x_3}$ from his preference list. Similarly, the remaining patients $p_1^{x_3}$, $p_4^{x_3}$, and $p_2^{x_3}$ are selected in the presented order.

 The final allocation of $patient-doctor$ pair is shown in Figure \ref{fig:gull2}.
\section{Experimental findings}
In this section, we compare the efficacy of the proposed mechanisms via simulations. The experiments are carried out in this section to provide a simulation based on the data (the strict preference ordering of the patients) generated randomly using the Random library in Python. Our proposed naive mechanism $i.e.$ RanPAM is considered as a benchmark scheme and is compared with TOAM (in case of full preferences) and TOAM-IComP (in case of incomplete preferences). 
\subsection{Simulation set-up}
For creating a real world healthcare scenario we have considered 10 different categories of patients and doctors for our simulation purpose. It is to be noted that, in each of the categories, some fixed number of patients and fixed number of doctors are present for taking consultancy and for providing consultancy respectively. \\
One of the scenario that is taken into consideration for simulation purpose is, say there are equal number of patients and doctors present in each of the categories under consideration along with the assumption that each of the patients are providing strict preference ordering (generated randomly) over all the available doctors in the respective categories. This scenario is referred as \emph{Scenario-1} in the rest of the paper.
\begin{table*}
\centering
 \caption{Simulation data set utilized for different scenarios}
\label{data}
\renewcommand{\arraystretch}{1.3}
\scalebox{0.6}{
\begin{tabular}{ l | l | l  l l l l l l l  l  l |l  l l l l l l l  l  l | l | l}
\hline

Scenarios & Preference profiles type ($\lesseqgtr$)  & \multicolumn{10}{|c|}{Number of Doctors (m)} & \multicolumn{10}{|c|}{Number of Patients (n)} & Total Doctors & Total Patients \\
&&$x_1$&$x_2$&$x_3$&$x_4$&$x_5$&$x_6$&$x_7$&$x_8$&$x_9$&$x_{10}$&$x_1$&$x_2$&$x_3$&$x_4$&$x_5$&$x_6$&$x_7$&$x_8$&$x_9$&$x_{10}$&&\\
\hline
\emph{Scenario-1}&Full preference ($m=n$)&10&10&10&10&10&10&10&10&10&10&10&10&10&10&10&10&10&10&10&10&100&100\\
&&20&20&20&20&20&20&20&20&20&20&20&20&20&20&20&20&20&20&20&20&200&200\\
&&30&30&30&30&30&30&30&30&30&30&30&30&30&30&30&30&30&30&30&30&300&300\\
&&40&40&40&40&40&40&40&40&40&40&40&40&40&40&40&40&40&40&40&40&400&400\\
&&50&50&50&50&50&50&50&50&50&50&50&50&50&50&50&50&50&50&50&50&500&500\\
\hline
\emph{Scenario-2}&Partial preference ($m=n$)&10&10&10&10&10&10&10&10&10&10&10&10&10&10&10&10&10&10&10&10&100&100\\
&&20&20&20&20&20&20&20&20&20&20&20&20&20&20&20&20&20&20&20&20&200&200\\
&&30&30&30&30&30&30&30&30&30&30&30&30&30&30&30&30&30&30&30&30&300&300\\
&&40&40&40&40&40&40&40&40&40&40&40&40&40&40&40&40&40&40&40&40&400&400\\
&&50&50&50&50&50&50&50&50&50&50&50&50&50&50&50&50&50&50&50&50&500&500\\
\hline
\emph{Scenario-3}&Partial preference ($m>n$)&12&9&8&13&10&7&10&11&11&9&8&6&6&10&8&5&9&7&8&8&100&75\\
&&20&15&27&21&17&19&23&21&23&14&14&10&21&17&13&15&20&16&18&11&200&155\\
&&27&27&33&36&33&24&39&21&31&29&19&21&25&22&29&20&17&14&25&28&300&220\\
&&40&41&45&41&44&32&38&36&40&43&37&34&43&37&38&27&36&32&36&40&400&360\\
&&45&55&55&50&35&50&65&40&45&60&43&53&47&45&30&49&60&35&44&56&500&462\\
\hline
\emph{Scenario-4}&Partial preference ($m<n$)&8&6&6&10&8&5&9&7&8&8&12&9&8&13&10&7&10&11&11&9&75&100\\
&&14&10&21&17&13&15&20&16&18&11&20&15&27&21&17&19&23&21&23&14&155&200\\
&&19&21&25&22&29&20&17&14&25&28&27&27&33&36&33&24&39&21&31&29&220&300\\
&&37&34&43&37&38&27&36&32&36&40&40&41&45&41&44&32&38&36&40&43&360&400\\
&&43&53&47&45&30&49&60&35&44&56&45&55&55&50&35&50&65&40&45&60&462&500\\
\hline
\end{tabular}
}
\end{table*}

Next, the more general scenario with equal number of patients and doctors in each of the categories can be obtained by relaxing the constraint that all the available patients are providing the strict preference ordering over all the available doctors in categories under consideration. Here, it may be the case that, in each of the categories some subset of the patients are providing the strict preference ordering over the subset of the available doctors. This scenario is referred as \emph{Scenario-2} in the rest of the paper.\\
In the series of different scenarios, next we have considered the utmost general set-up where there are \emph{n} number of patients and \emph{m} number of doctors such that $m \neq n$ ($m>n$ and $m<n$). In this, the subset of patients are providing the strict preference ordering over the subset of the available doctors in each categories under consideration. The scenario with $m>n$ is referred as \emph{scenario-3} and the scenario with $m<n$ is referred as \emph{scenario-4} in the future references. The data that is utilized for the simulation purpose in all the four scenarios are shown in table \ref{data}.

\subsection{Performance metrics}
The performance of the proposed mechanisms is measured under the banner of two important parameters:\\
$\bullet$ \textbf{Efficiency loss (EL).} It is the sum of the difference between the index of the doctor allocated from the agent preference list to the index of the most preferred doctor by the agent from his preference list. Mathematically, the $EL$ is defined as: EL = $\sum_{i=1}^{n} (\overline{I}_{i_{\mathcal{A}}} - \overline{I}_{i_{\mathcal{M}\mathcal{P}}})$
 where, $\overline{I}_{i_{\mathcal{A}}}$ is the index of the doctor allocated from the initially provided preference list of the patient \emph{i}, $\overline{I}_{i_{\mathcal{M}\mathcal{P}}}$ is the index of the most preferred doctor in the initially provided preference list of patient \emph{i}.
 Considering the overall available categories $x=\{x_1, x_2, \ldots, x_k\}$, the total efficiency loss (TEL) of the system is given as:
 \begin{equation}\label{eq:1}
 TEL = \sum_{x}\sum_{i=1}^{n} (\overline{I}_{i_{\mathcal{A}}} - \overline{I}_{i_{\mathcal{M}\mathcal{P}}})
 \end{equation} 
$\bullet$ \textbf{Number of best allocation (NBA).} It measures the number of patients (say $k$) gets their best choice (most preferred doctor) from their provided preference list over the available number of doctors. It is the sum of the number of agents getting their most preferred doctor from their provided preference list.  
\begin{figure*}
\begin{multicols}{2}
     \begin{subfigure}[b]{0.50\textwidth}
              \rotatebox{-90}{\includegraphics[scale=0.3]{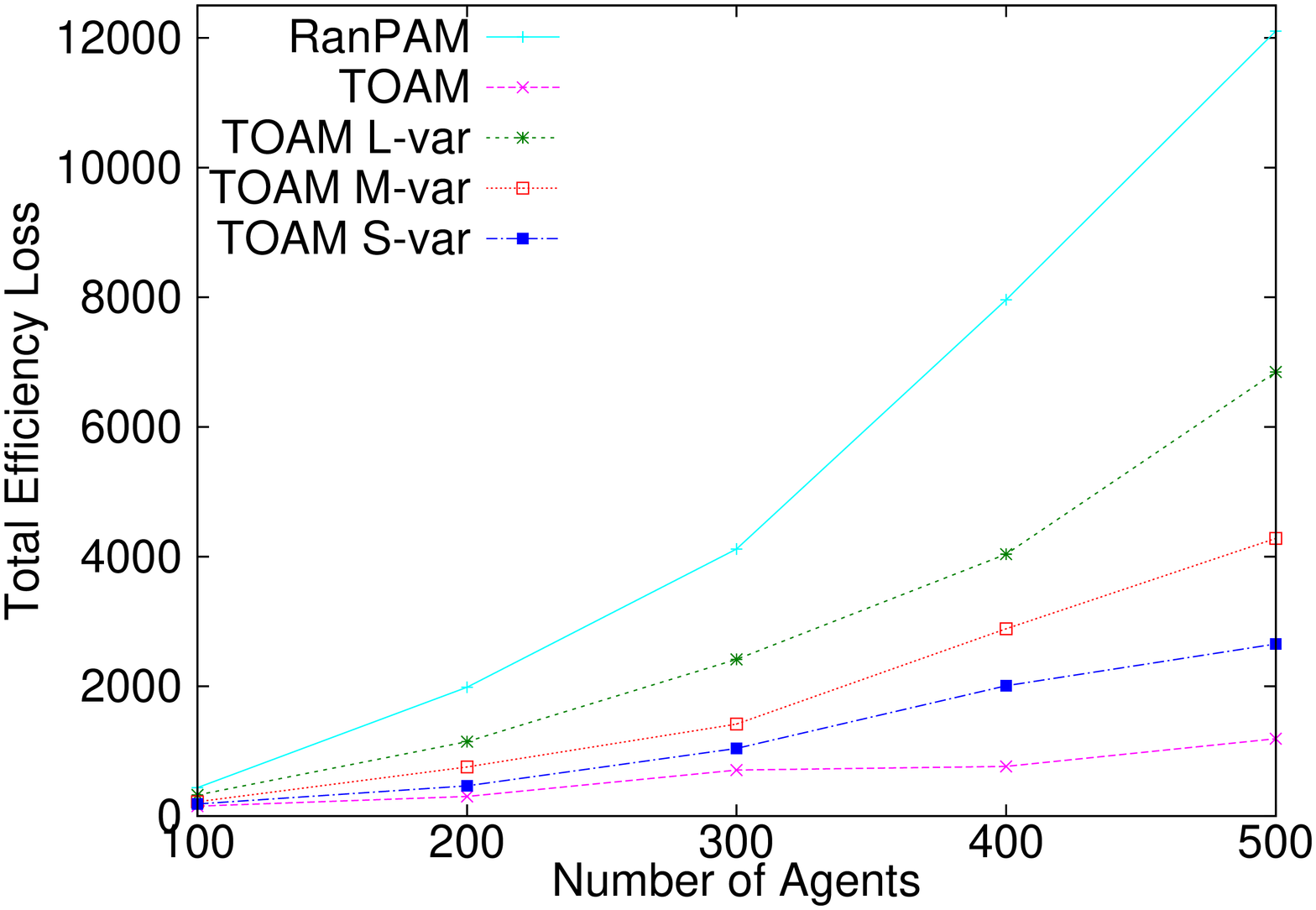}}
                \caption{Total Efficiency Loss (Scenario 1) }
                \label{fig:sce1}
        \end{subfigure}%
        \begin{subfigure}[b]{0.50\textwidth}
                \rotatebox{-90}{\includegraphics[scale=0.3]{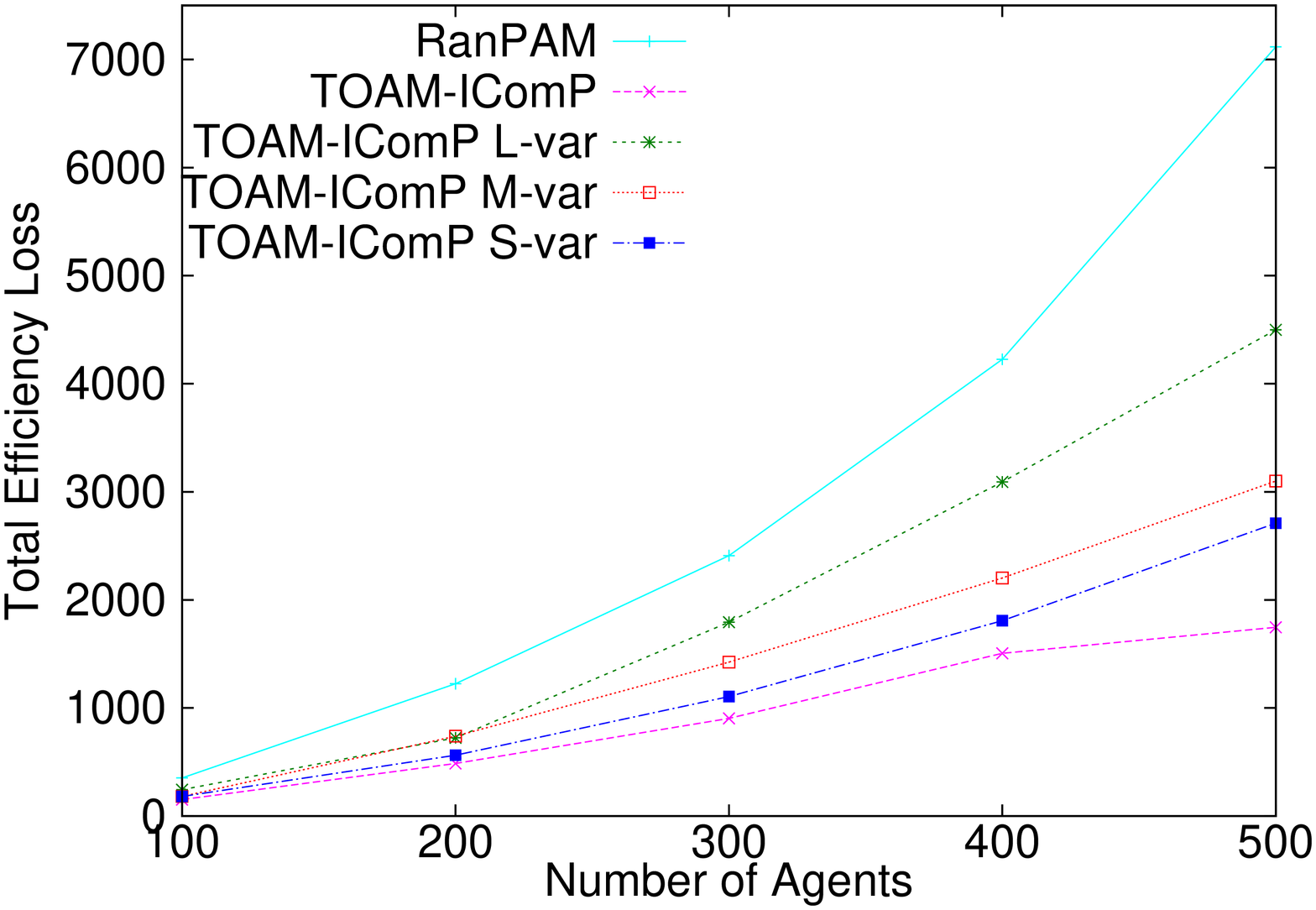}}
                \caption{Total Efficiency Loss (Scenario 2)}
                \label{fig:sce2}
        \end{subfigure}
        \vspace{10mm}
          \begin{subfigure}[b]{0.50\textwidth}
              \rotatebox{-90}{\includegraphics[scale=0.3]{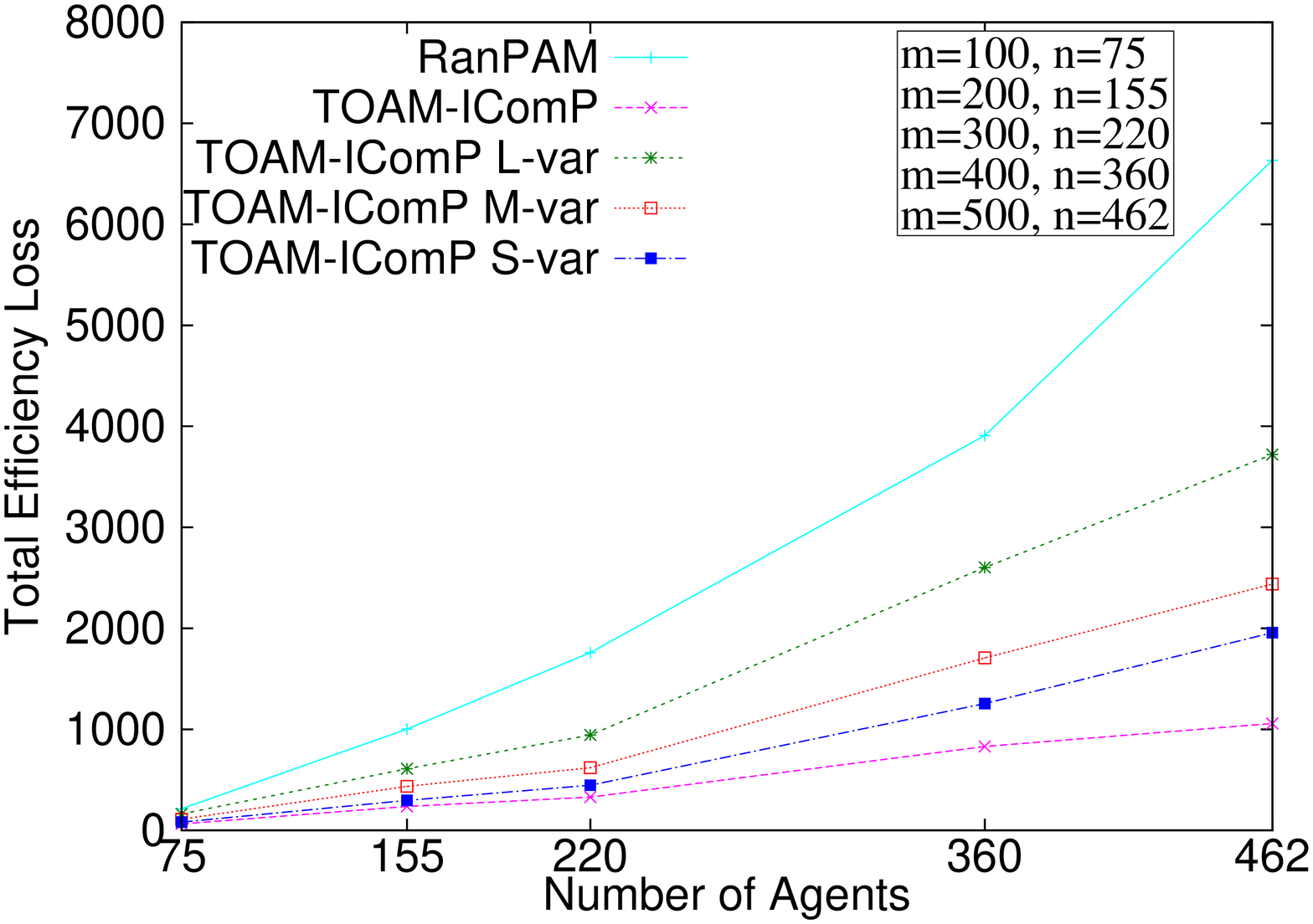}}
                \caption{Total Efficiency Loss (Scenario 3) }
                \label{fig:sce3}
        \end{subfigure}%
        \begin{subfigure}[b]{0.50\textwidth}
                \rotatebox{-90}{\includegraphics[scale=0.3]{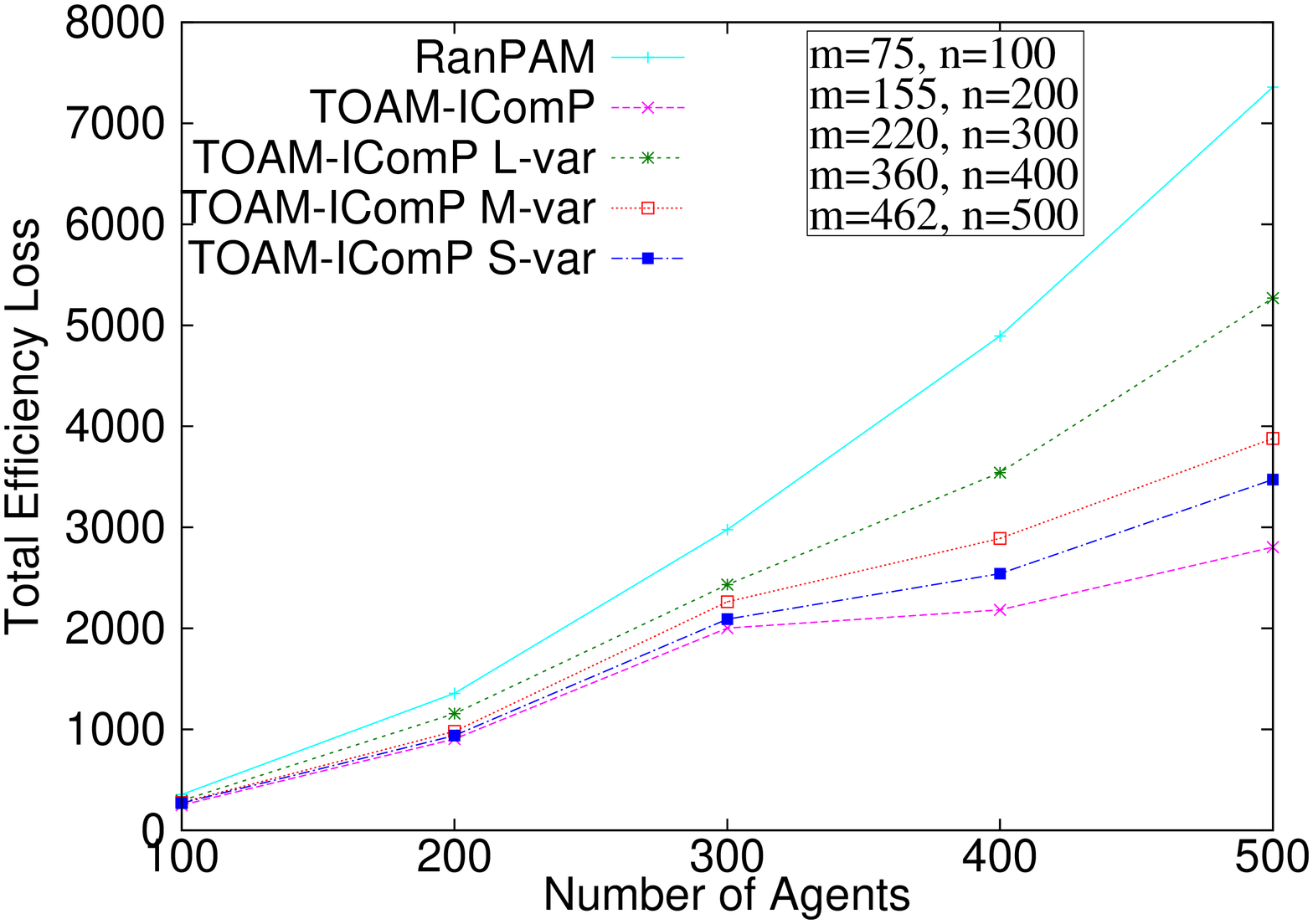}}
                \caption{Total Efficiency Loss (Scenario 4)}
                \label{fig:sce4}
        \end{subfigure}
      
\end{multicols}
  \caption{Total efficiency loss for different scenarios}\label{fig:animals1}
\label{fig:5}
\end{figure*}

\begin{figure*}
\begin{multicols}{2}
     \begin{subfigure}[b]{0.50\textwidth}
           \includegraphics[scale=0.63]{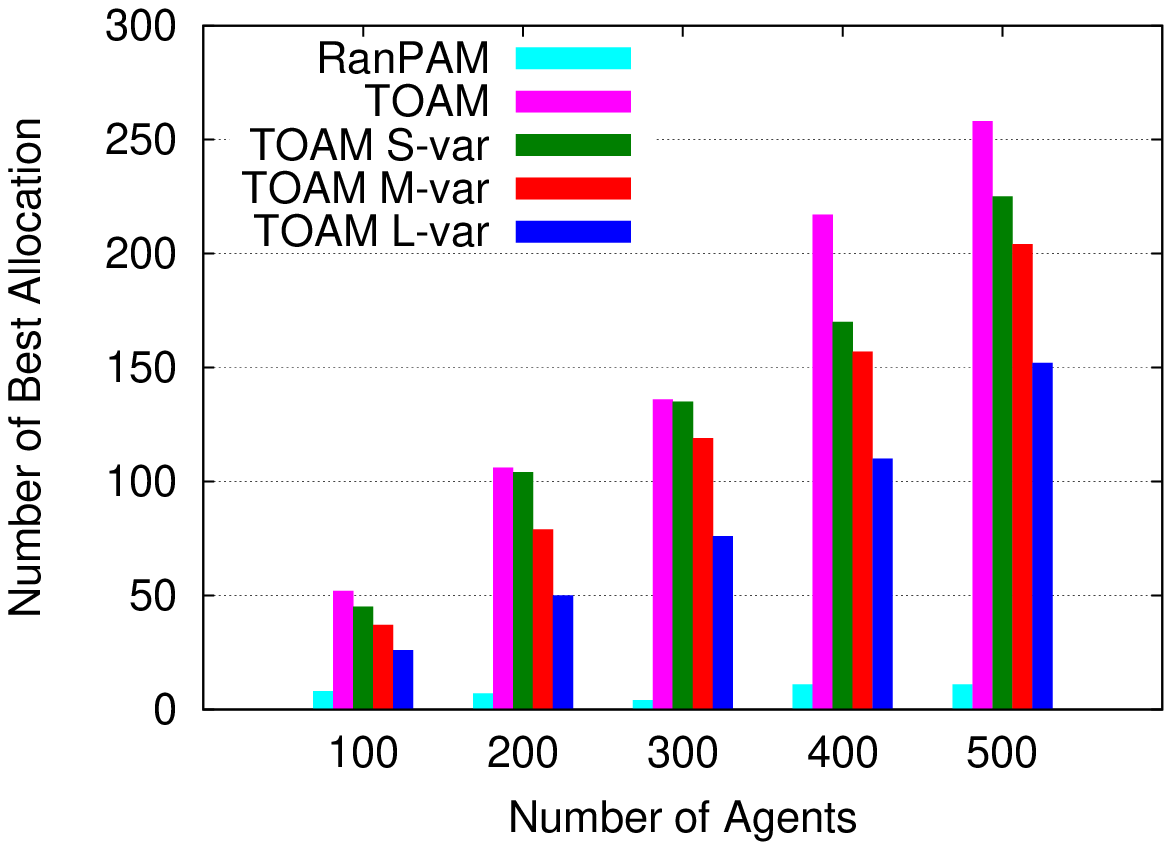}
                \caption{Total Number of Best Allocation (Scenario 1) }
                \label{fig:sce1b}
        \end{subfigure}%
        \begin{subfigure}[b]{0.50\textwidth}
        \includegraphics[scale=0.63]{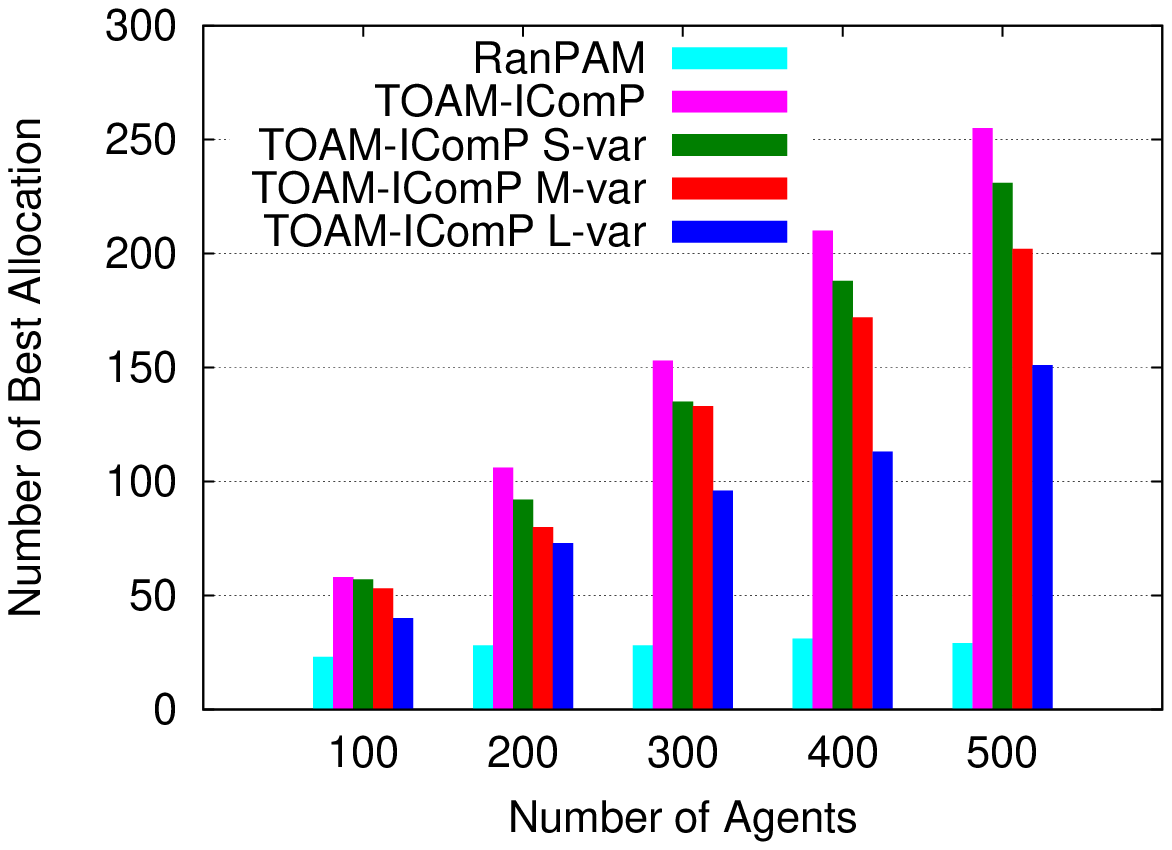}
                \caption{Total Number of Best Allocation (Scenario 2)}
                \label{fig:sce2b}
        \end{subfigure}
\vspace{10mm}
        \begin{subfigure}[b]{0.50\textwidth}
           \includegraphics[scale=0.63]{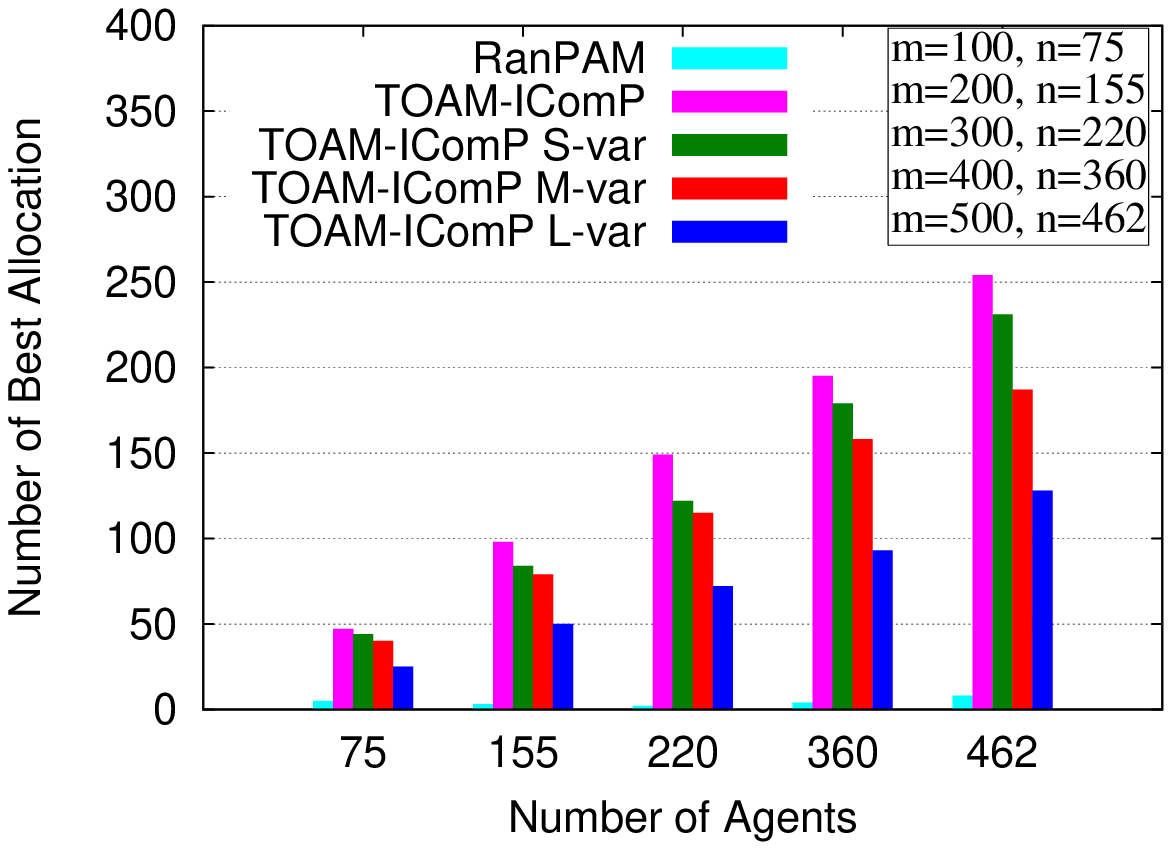}
                \caption{Total Number of Best Allocation (Scenario 3) }
                \label{fig:sce3b}
        \end{subfigure}%
        \begin{subfigure}[b]{0.50\textwidth}
        \includegraphics[scale=0.63]{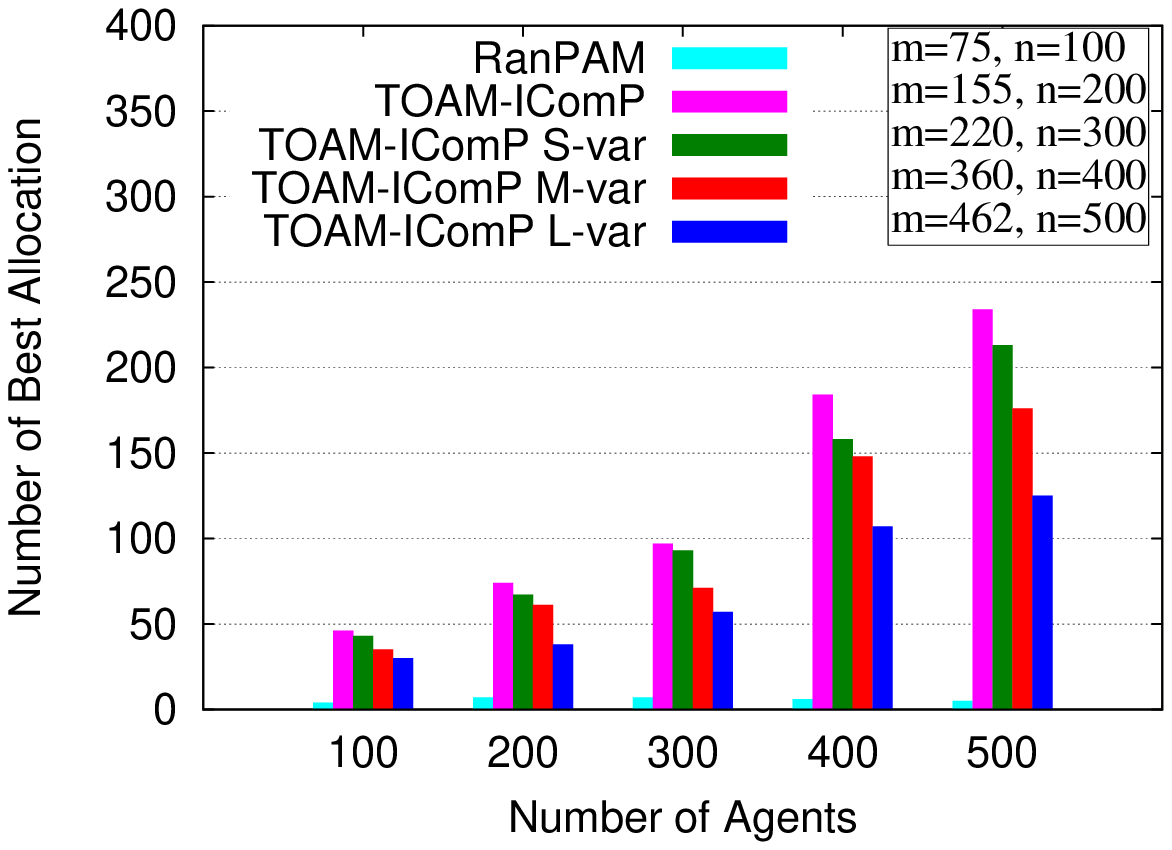}
                \caption{Total Number of Best Allocation (Scenario 4)}
                \label{fig:sce4b}
        \end{subfigure}
      
\end{multicols}
  \caption{Number of best allocation for different scenarios}\label{fig:animals1}
\label{fig:5}
\end{figure*}

\subsection{Simulation directions} 
In order to analyse the effect of manipulative behaviour of the agents on the proposed optimal mechanisms ($i.e.$ TOAM and TOAM-IComP), the two proposed directions are considered:
(1) When all the agents are reporting their true preference list.
(2) When subset of the agents are misreporting their true preference list.

\subsection{Result analysis}
 In this section, the result is simulated following the directions mentioned in Subsection C.
As the patients are varying their true preference list, the next question that comes is that, how many of the patients can vary their true preference list ($i.e.$ what fraction of the total available patients can vary their true preference list?). To answer this question, the calculation is done using indicator random variable.\\ 
$-$\textbf{Expected amount of variation}
The following analysis mathematically justifies the idea of choosing the parameters of variation.
The analysis is motivated
by \cite{Coreman_2009}. Let $\mathcal{N}_{i}$ be the random variable associated with the event in which $i^{th}$ patient varies his true preference ordering.
Thus, $\mathcal{N}_{i}$ = \{$i^{th}$ patient varies preference ordering\}. We have, from the definition of expectation that $E[\mathcal{N}_{i}]$ = Pr$\{i^{th}$ \emph{patient varies preference ordering}$\}$.
Let $\mathcal{N}$ be the random variable denoting the total number of patients vary their preference ordering. By using the properties of random variable, it can be written that 
$\mathcal{N} = \sum_{i =1}^{n} \mathcal{N}_i$. We wish to compute the expected number of variations, and so we take the expectation both sides and 
by linearity of expectation we can write $E[\mathcal{N}]$ = $\sum_{i =1}^{n} E[\mathcal{N}_i]$ = $\sum_{i=1}^{n}$(Pr$\{i^{th}$ 
\emph{patient varies preference ordering}$\}$) = $\sum_{i=1}^{n} 1/8$ = $n/8$. Here, 
Pr$\{i^{th}$ \emph{patient varies preference ordering}$\}$ is the probability that given a patient whether he will vary his 
true preference ordering. The probability of that is taken as $1/8$ (small variation). If the number of agents varies from $1/4$ and $1/2$, then the expected number of patient that may vary their preference ordering 
can be $n/4$ (medium variation), and $n/2$ (large variation) respectively.\\
\noindent \newline In Fig. \ref{fig:sce1} and Fig. \ref{fig:sce2}-\ref{fig:sce4}, it can be seen that the total efficiency loss of the system in case of RanPAM is more than the total efficiency loss of the system in case of TOAM and TOAM-IComP respectively. This is due to the fact that, dissimilar to the RanPAM, TOAM and TOAM-IComP allocates the best possible doctors to the patients from their revealed preference list. Due to this reason, the value returned by equation \ref{eq:1} in case of TOAM and TOAM-IComP is very small as compared to RanPAM. In Fig. \ref{fig:sce1}, when the agents are varying (misreporting) their true preference ordering, then the TEL of the patients in case of TOAM with large variation (TOAM L-var) is more than the TEL in TOAM with medium variation (TOAM M-var) is more than the TEL in TOAM with small variation (TOAM S-var) is more than the TEL in TOAM without variation. As it is natural from the construction of the TOAM.\\ 
Considering the case of incomplete preferences in Fig. \ref{fig:sce2}-\ref{fig:sce4}, when the subset of agents are varying their true preference ordering, then the TEL of the patients in case of TOAM-IComP with large variation (TOAM-IComP L-var) is more than the TEL in TOAM-IComP with medium variation (TOAM-IComP M-var) is more than the TEL in TOAM-IComP with small variation (TOAM-IComP S-var) is more than the TEL in TOAM-IComP without variation. As this is evident from the construction of the TOAM-IComP.\\
Considering the case of our second parameter $i.e$ NBA, in Fig. \ref{fig:sce1b} and Fig. \ref{fig:sce2b}-\ref{fig:sce4b}, it can be seen that the NBA of the system in case of RanPAM is less than the NBA of the system in case of TOAM and TOAM-IComP respectively. This is due to the fact that, dissimilar to the RanPAM, TOAM and TOAM-IComP allocates the best possible doctors to the patients from their preference list. In Fig. \ref{fig:sce1b}, when the agents are varying their true preference ordering, then the NBA of the patients in case of TOAM with large variation (TOAM L-var) is less than the NBA in TOAM with medium variation (TOAM M-var) is less than the NBA in TOAM with small variation (TOAM S-var) is less than the NBA in TOAM without variation. As it is natural from the construction of the TOAM.\\ 
In Fig. \ref{fig:sce2b}-\ref{fig:sce4b}, when the subset of agents are varying their true preference ordering, then the NBA of the patients in case of TOAM-IComP with large variation (TOAM-IComP L-var) is less than the NBA in TOAM-IComP with medium variation (TOAM-IComP M-var) is less than the NBA in TOAM-IComP with small variation (TOAM-IComP S-var) is less than the TEL in TOAM-IComP without variation. As this is evident from the construction of the TOAM-IComP.
\section{Conclusions and future works}
In this paper, we studied the problem of hiring renowned expert consultants (doctors) from around the world, to serve BIG patients of our society under \emph{zero} budget environment. The work can be further extended to many different settings, e.g. multiple doctors are allocated to a BIG patient in different hospitals around the globe. The other interesting direction is studying the discussed set-ups under budget constraints environment.  

\section*{Acknowledgement}
We would like to thank Prof. Y. Narahari and members of the Game Theory Lab. at IISc Bangalore for their useful advices. We would like to thank the faculty members, and PhD research scholars of the department for their valuable suggestions. We highly acknowledge the effort undertaken by Ministry of Electronics and Information Technology Government of India, Media Lab Asia, and Government of India Ministry of Human Resource Development (MHRD).

\balance
\bibliographystyle{unsrt}
\bibliography{phd}

\begin{thebibliography}{10}

\bibitem{Zhao2014CCC}
D.~Zhao, X.Y. Li, and Ma. Huadong.
\newblock How to crowdsource tasks truthfully without sacrificing utility:
  Online incentive mechanisms with budget constraint.
\newblock In {\em Proceeding of Annual International Conference on Computer
  Communications}, pages 1213--1221, Toronto, Canada, 2014.

\bibitem{Yang2012MCN}
D.~Yang, G.~Xue, X.~Fang, and J.~Tang.
\newblock Crowdsourcing to smart-phones: Incentive mechanism design for mobile
  phone sensing.
\newblock In {\em Proceeding of 18th Annual International Conference on Mobile
  Computing and Networking}, pages 173--184, Istanbul, Turkey, 2012. ACM.

\bibitem{Lee2010PMC}
J.~S. Lee and B.~Hoh.
\newblock Dynamic pricing incentive for participatory sensing.
\newblock {\em Elsevier Journal of Pervasive and Mobile Computing},
  6(6):693--708, 2010.

\bibitem{Alshurafa_2015}
N.~Alshurafa, J.~A. Eastwood, S.~Nyamathi, J.~J. Liu, W.~Xu, H.~Ghasemzadeh,
  M.~Pourhomayoun, and M.~Sarrafzadeh.
\newblock Improving compliance in remote healthcare systems through smartphone
  battery optimization.
\newblock {\em IEEE Journal of Biomedical and Health Informatics},
  19(1):57--63, 2015.

\bibitem{Kau_bhi}
L.~J. Kau and C.~S. Chen.
\newblock A smart phone-based pocket fall accident detection, positioning, and
  rescue system.
\newblock {\em IEEE Journal of Biomedical and Health Informatics},
  19(1):44--56, 2015.

\bibitem{Berrada_1996}
I.~Berrada, J.A. Ferland, and P.~Michelon.
\newblock A multi-objective approach to nurse scheduling with both hard and
  soft constraints.
\newblock {\em Socio-Economic Planning Sciences}, 30(3):183--193, 1996.

\bibitem{Pierskalla_1994}
W.~P. Pierskalla and D.~J. Brailer.
\newblock Applications of operations research in health care delivery.
\newblock {\em Handbooks in Operations Research and Management Science}, 6,
  1994.

\bibitem{Weil_1995}
G.~Weil, K.~Heus, P.~Francois, and M.~Poujade.
\newblock Constraint programming for nurse scheduling.
\newblock In {\em Engineering in Medicine and Biology}, pages 417--422, 1995.

\bibitem{Worthington_1988}
D.~Worthington and M.~Guy.
\newblock Allocating nursing staff to hospital wards: A case study.
\newblock {\em European Journal of Operational Research}, 33(2):174--182, 1988.

\bibitem{Carter_2001}
M.~W. Carter and S.~D. Lapiere.
\newblock Scheduling emergency room physicians.
\newblock {\em Health Care Management Science}, 4(4):347--360, 2001.

\bibitem{Vassilacopoulos_1985}
G.~Vassilacopoulos.
\newblock Allocating doctors to shifts in an accident and emergency department.
\newblock {\em Journal of the Operational Research Society}, 36(6):517--523,
  1985.

\bibitem{Beaulieu_2000}
H.~Beaulieu, J.~A. Ferland, B.~Gendron, and P.~Michelon.
\newblock A mathematical programming approach for scheduling physicians in the
  emergency room.
\newblock {\em Health Care Management Science}, 3(3):193--200, 2000.

\bibitem{Wang_2007}
C.~W. Wang, L.~M. Sun, M.~H. Jin, C.~J. Fu, L.~Liu, C.~H. Chan, and C.~Y. Kao.
\newblock A genetic algorithm for resident physician scheduling problem.
\newblock In {\em Proceedings of the $9^{th}$ Annual Conference on Genetic and
  Evolutionary Computation}, pages 2203--2210, New York, USA, 2007.

\bibitem{Software_1}
American college of emergency physicians.
\newblock In {\em Directory of software in emergency medicine}, Dallas, 1998.
  (http://www.acep.org).

\bibitem{Software_2}
Bytebloc software, epsked 3.0.
\newblock In {\em ByteBloc Software}, LongBeach, 1995.
  (http://www.bytebloc.com).

\bibitem{Software_3}
Msi software, physician scheduler 3.2.
\newblock In {\em MSI Software}, Fairfax, 1998. (http://www.msisoftware.com).

\bibitem{Software_4}
Peake software laboratories: Tangier emergency physician scheduling software.
\newblock In {\em Peake Software}. (http://peakesoftware.com/peake/).

\bibitem{TLHP_INFO_2014}
T.~Luo, H.~P. Tan, and L.~Xia.
\newblock Profit-maximizing incentive for participatory sensing.
\newblock In {\em IEEE International conference on Computer Communications
  (INFOCOM)}, pages 127--135, April-May 2014.

\bibitem{cardoen_b_2010}
B.~Cardoen, E.~Demeulemeester, and J.~Belien.
\newblock Operating room planning and scheduling: A literature review.
\newblock {\em European journal of operational research}, 201(3):921--932,
  2010.

\bibitem{Magerlein_JM_1978}
J.~M. Magerlein and J.~B. Martin.
\newblock Surgical demand scheduling: A review.
\newblock {\em Health service research}, 13(4):418--433, 1978.

\bibitem{Blake_J_1997}
J.~Blake and M.~Carter.
\newblock Surgical process scheduling.
\newblock {\em Journal of the society for health systems}, 5(3):17--30, 1997.

\bibitem{Vikash2015EAS}
V.~K. Singh, S.~Mukhopadhyay, N.~Debnath, and A.M. Chowdary.
\newblock Auction aware selection of doctors in e-healthcare.
\newblock In {\em Proceeding of $17^{th}$ Annual International Conference on
  E-health Networking, Application and services (HealthCom)}, pages 363--368,
  Boston, USA, 2015.

\bibitem{Gale_AMM_1962}
D.~Gale and L.S. Shapley.
\newblock College admissions and the stability of marriage.
\newblock {\em The American Mathematical Monthly}, 69(1):9--15, 1962.

\bibitem{Shapley_2013}
L.~Shapley and H.~Scarf.
\newblock On cores and indivisibility.
\newblock {\em Journal of Mathematical Economics}, 1:23--37, March 1974.

\bibitem{NNis_Pre_2007}
N.~Nisan, T.~Roughgarden, E.~Tardos, and V.~V. Vazirani.
\newblock {\em Algorithmic Game Theory}.
\newblock Cambridge University Press, 2007.

\bibitem{Roughgarden_T_2013}
T.~Roughgarden.
\newblock Cs364a: Algorithmic game theory, lecture \#9: Beyond quasi-linearity,
  October 2013.

\bibitem{T.roughgarden_2016}
T.~Roughgarden.
\newblock Cs269i: Incentives in computer science, 2016.
\newblock Lecture \#1: The Draw and College Admissions.

\bibitem{Manlove_2016}
B.~Klaus, D.~F. Manlove, and F.~Rossi.
\newblock {\em Matching under preferences}.
\newblock 2016.

\bibitem{Coreman_2009}
T.~H. Cormen, C.~E. Leiserson, R.~L. Rivest, and C.~Stein.
\newblock {\em Introduction to Algorithms}.
\newblock McGraw-Hill Higher Education, 2nd edition, 2001.

\bibitem{Gries_2009}
D.~Gries.
\newblock {\em The Science of Programming}.
\newblock Springer-Verlag New York, Inc., Secaucus, NJ, USA, 1st edition, 1987.

\end{thebibliography}

\vfill


\end{document}